\documentclass[a4paper]{article}
\usepackage[normalem]{ulem}
\usepackage{titlesec}
\usepackage{comment}
\usepackage[english]{babel}
\usepackage[T1]{fontenc}
\usepackage{amssymb}
\usepackage{amsthm}
\usepackage[a4paper,top=2.5cm,bottom=2.5cm,left=2cm,right=2cm,marginparwidth=1.75cm]{geometry}
\usepackage{enumerate}
\usepackage{wrapfig}
\usepackage{enumerate}
\usepackage{enumitem}
\def\R{{\cal R}}
\usepackage[dvipsnames]{xcolor}
\usepackage{lineno}

\newcommand{\diag}{\operatorname{diag}}

\newcommand{\Ker}{\operatorname{Ker}}
\newcommand{\rank}{\operatorname{rank}}
\newcommand{\range}{\operatorname{Range}}
\newcommand{\dg}{\operatorname{diag}}
\newcommand{\T}{\operatorname{\,T}}

\usepackage[export]{adjustbox}
\usepackage{subfig}
\usepackage{amsmath}
\usepackage{graphicx}
\usepackage[colorinlistoftodos]{todonotes}
\usepackage[colorlinks=true, allcolors=blue,hypertexnames=false]{hyperref}
\usepackage{float}

\usepackage{amsmath,amsfonts,amssymb,amsthm}
\usepackage{mathtools}
\usepackage{commath}
  
\let\norm\undefined 
\DeclarePairedDelimiter\norm{\lVert}{\rVert}

\title{An adaptation of InfoMap to absorbing random walks using absorption-scaled graphs}
\author{Esteban Vargas, Mason A. Porter,  Joseph H. Tien}

\newtheorem{definition}{Definition}
\newtheorem{proposition}{Proposition}
\newtheorem*{proposition*}{Proposition}
\newtheorem*{remark*}{Remark}

\usepackage{algorithmic}            
\usepackage{algorithm}

\titleformat{\paragraph}
{\normalfont\normalsize\bfseries}{\theparagraph}{1em}{}
\titlespacing*{\paragraph}
{0pt}{3.25ex plus 1ex minus .2ex}{1.5ex plus .2ex}

\newcounter{savealgorithm}
\newenvironment{subalgorithms}
 {%
  \stepcounter{algorithm}%
  \edef\currentthealgorithm{\thealgorithm}%
  \setcounter{savealgorithm}{\value{algorithm}}%
  \setcounter{algorithm}{0}%
  \renewcommand{\thealgorithm}{\currentthealgorithm\alph{algorithm}}%
 }
 {%
  \setcounter{algorithm}{\value{savealgorithm}}%
 }

\begin{document}
\maketitle

\begin{abstract}

InfoMap is a popular approach to detect densely connected ``communities'' of nodes in networks. To detect such communities, InfoMap uses random walks and ideas from information theory. 
Motivated by the dynamics of disease spread on networks, whose nodes can have heterogeneous disease-removal rates, we adapt InfoMap to absorbing random walks. To do this, we use {\it absorption-scaled graphs} (in which edge weights are scaled according to absorption rates) and Markov time sweeping. One of our adaptations of InfoMap converges to the standard version of InfoMap in the limit in which the node-absorption rates approach $0$. We demonstrate that the community structure that one obtains using our adaptations of InfoMap can differ markedly from the community structure that one detects using methods that do not account for node-absorption rates. We also illustrate that the community structure that is induced by heterogeneous absorption rates can have important implications for susceptible--infected--recovered (SIR) dynamics on ring-lattice networks. For example, in some situations, the outbreak duration is maximized when a moderate number of nodes have large node-absorption rates. 

\end{abstract}


\section{Introduction}

Random walks are one of the most fundamental dynamical processes, and many studies have used random walks on networks (i.e., graphs) to gain insights into network structure and how such structure affects dynamical processes \cite{masuda2017random}. Much research has focused on standard random walks, in which the distribution of the occupation probabilities of a network's nodes converges to a stationary distribution with all positive entries in the limit of infinitely many walker steps. It is important to understand the relationship between network structure and different types of random walks. {In the present paper, we consider absorbing random walks, in which the probability to reach an ``absorbing state'' converges to $1$ as the number of walker steps becomes infinite. We examine dynamical processes that involve absorbing random walks, for which there is a nonzero rate (the so-called ``absorption rate'') of transitioning to an absorbing state from each node of a graph.

Absorbing random walks have been used to develop centrality measures \cite{gurfinkel2020absorbing}, other methods to rank the nodes of a network \cite{zhu2007improving}, transduction algorithms (which one can use to infer the labels of the nodes of a graph from the labels of a subset of the nodes) \cite{de2017transduction}, and more. For example, Jaydeep et al.~\cite{de2017transduction} proposed a transduction algorithm that uses the number of visits before absorption of an absorbing random walk as a measure of affinity between the nodes of a graph. Absorbing random walks also arise naturally in many modeling contexts, including population dynamics \cite{fletcher2019towards}, the spread of infectious diseases on networks \cite{kim2022}, and the propagation of content in online social networks \cite{bhagat2011node}. In the setting of population dynamics, consider a collection of habitat patches that are connected through some mobility network. In this context, a random walk corresponds to an individual moving between patches and absorption corresponds to death \cite{acevedo2015conservation,fletcher2023framework, fletcher2019towards}. 
Similar situations occur in contagion processes, where one can model the transfer of a pathogen between sites as a random walk and pathogen clearance (e.g., recovery from infection) as absorption. This interpretation of pathogen transfer underlies the next-generation-matrix approach \cite{vandendriessche2002} to calculate the basic reproduction number $\R_0$, which is a staple quantity of mathematical epidemiology that indicates the mean number of secondary infections that arise when a single infected individual enters a population of susceptible individuals \cite{Brauer2019}.

In the present paper, we examine absorbing random walks on graphs in which different nodes can have different absorption rates, inducing an ``effective'' network structure that is reflected only partially by the edge weights of a network. Many notions of network community structure arise from the analysis of random walks \cite{fortunato2016community,masuda2017random}, and we expect different types of random walks to yield different community structures \cite{jeub2015,jeub2017}. A ``community'' in a network is a tightly knit set of nodes that is connected sparsely to other tightly knit sets of nodes~\cite{fortunato2016community,Porter2009}. Communities are a common feature of many real-world networks, and community structure influences dynamical processes such as the spread of infectious diseases \cite{stegehuis2016} and online content \cite{huang2006,weng2013}. For example, community structure can affect the size and duration of a disease outbreak~\cite{salathe2010}. There is intense interest in understanding how community structure and node characteristics combine to influence contagions on networks \cite{levy2022,ratmann2020,ruktanonchai2016b}.

We develop community-detection algorithms that account for node-absorption rates. We adapt the widely-used community-detection algorithm InfoMap \cite{rosvall2009map,rosvall2008maps,infomap2023} to absorbing random walks and thereby account for heterogeneous node-absorption rates in the detected communities. In our adaptation, we apply InfoMap to \emph{absorption-scaled graphs}, which account for absorption by scaling the edge weights of a network \cite{jacobsen2018generalized}. These absorption-scaled graphs are related to their associated absorbing random walks by a generalized inverse (the so-called ``absorption inverse'') and a fundamental matrix \cite{jacobsen2018generalized}. We use absorption inverses and results from \cite{jacobsen2018generalized} to study the absorption-scaled graphs that are associated with our adaptations of InfoMap.

Community structure can greatly influence disease dynamics on networks \cite{mistry2021inferring,pastor2015epidemic}. Salath\'{e} and Jones \cite{salathe2010} illustrated that changes in community structure are correlated with changes in disease quantities for susceptible--infected--recovered (SIR) dynamics on a network. One of their findings is that outbreak duration can achieve a maximum at intermediate modularity values. Inspired by Salath\'{e} and Jones \cite{salathe2010}, we use our adaptation of InfoMap to study an example contagion process that illustrates how absorption in disease dynamics affects community structure, which in turn affects disease spread. We investigate the association between changes in the effective community structure that is induced by the node-absorption rates with disease quantities such as outbreak size and duration.

Our paper proceeds as follows. In Section \ref{sec:MapField}, we present the original InfoMap algorithm and the Markov time-sweeping technique that we use in our adaptations of InfoMap. In Section \ref{sec:scaledGrahExtension}, we apply InfoMap to absorption-scaled graphs. In Section \ref{sec:redefMap}, we introduce a definition of a map function $L^{(a)}$ for absorbing random walks. We relate this new map function to our adaptations of InfoMap. In Section \ref{sec:examples}, we discuss toy examples to illustrate that our adaptations of InfoMap yield effective community structures, which arise from the heterogeneous node-absorption rates. In Section \ref{sec:small_world}, we examine effective community structure and SIR disease dynamics on a synthetic network. In Section \ref{sec:discussion}, we summarize and discuss our key conclusions. In Appendix \ref{appendix}, we present the proofs of {three key} propositions from Section \ref{sec:redefMap}. Our code is available at \url{https://gitlab.com/esteban_vargas_bernal/extending-infomap-to-absorbing-random-walks}.

\section{Background on InfoMap}\label{sec:MapField}

We now present background material on InfoMap and an adaptation of it to continuous-time Markov chains using Markov time sweeping \cite{schaub2012encoding}. We summarize our notation in Table \ref{table:notation}. 
 
\begin{table}[H]
\begin{centering}
\begin{tabular}{lll}
{\bf Symbol} & {\bf Expression} & {\bf Description}  \\ \hline \hline
$G$ & & {D}irected, weighted graph  \\ \hline
$A$ & & {A}djacency matrix of a directed, weighted graph \\ \hline
$W$ & & {D}iagonal matrix of out-degrees  \\ \hline
$\mathcal{L}$ & $W - A$ & {U}nnormalized graph Laplacian matrix \\ \hline
$P$ & $AW^{-1}$ & {T}ransition-probability matrix of the Markov chain\\ 
& & that is associated with $A$ \\ \hline
$M$ & $\{M_1,\ldots,M_m\}$ & {P}artition of a graph into communities \\ \hline
$M_i$ & $\{k_1,\ldots,k_{{n_i}}\}$ & Community that consists of nodes $k_1,\ldots, {k_{n_i}}$ \\ \hline
{$\pi$} & {$(\pi_1,\ldots,\pi_{n})^{\T}$} & {Probability distribution on the set of nodes for the map function} 
\\ \hline
$q_{i\curvearrowright}$ & $\sum_{j \in M_i, k \notin M_i} \pi_j p_{kj}$ & Probability of a transition out of community $M_i$ \\ \hline 
$q_{i\curvearrowleft}$ & $\sum_{k \in M_i, j \notin M_i} \pi_j p_{kj}$  & Probability of a transition into community $M_i$ \\ \hline
$q_\curvearrowleft$ & $\sum_{i \in \{1,\ldots,m\}} q_{i\curvearrowleft}$ & Probability of a transition into a different community \\ \hline

{$p_\circlearrowright^i$} & {$ q_{i\curvearrowright } + \sum_{j\in M_i} \pi_j$} & {Normalization factor for the probability distribution $\mathcal{P}^i$} \\ \hline  

${\mathcal{Q}}$ & $\left(q_{1\curvearrowleft}/q_\curvearrowleft, \ldots, q_{m\curvearrowleft}/q_\curvearrowleft \right)^{\T}$ &  {Probability distribution for the map function that } \\
& & {is associated with transitions into a different community} \\ \hline

$\mathcal{P}^i $ & $ \left( q_{i \curvearrowright}/p_\circlearrowright^i, \pi_{k_1}/p_\circlearrowright^i, \ldots, \pi_{k_{{n_i}}}/p_\circlearrowright^i  \right)^{\T} $ & {Probability distribution  for the map function that } \\
& &  {is associated with transitions out of community $M_i$ } \\ \hline 

${\cal H}$ & ${\cal H(P)}, {\cal{H(Q)}}$ & {E}ntropy of a distribution\\ \hline
$L(M,P,\pi)$ & $q_\curvearrowleft \mathcal{H}(\mathcal{Q}) + \sum_{i=1}^m p_\circlearrowright ^i \mathcal{H}(\mathcal{P}^i)$ & {M}ap function \\ \hline
$H$ &  & {Diagonal matrix with non-negative entries on its diagonal} \\ \hline
$\vec{d}$ & & {G}eneric absorption-rate vector \\ \hline
$\vec{\delta}$ & & {N}ode-absorption-rate vector \\ \hline
$D_\delta$ & $\diag\{\vec{\delta}\}$ & {D}iagonal matrix of node-absorption rates \\  \hline
$\vec{d}_{\mathrm{s}}(D_\delta, H)$ & $(d_{\mathrm{s}})_i = h_i w_i + \delta_i$ & {S}caled rate vector, which is given by \\
& & the diagonal of the matrix $HW + D_\delta$\\ \hline
$\tilde{G}(D_\delta, H)$ & & {A}bsorption-scaled graph that is associated \\
& & with the pair $(G,\vec{d}_{\mathrm{s}}(D_\delta,H))$\\ \hline
$\tilde{\mathcal{L}}(D_\delta, H)$ &  $(W-A)(HW + D_{\delta})^{-1}$ & {U}nnormalized graph Laplacian matrix of an\\
& &absorption-scaled graph \\ \hline
$\vec{u}$ & &  {V}ector in $\Ker {\mathcal{L}}$  with non-negative entries \\
& & that satisfy $\sum_{i=1}^n u_i =1$ \\ \hline
$P_e(D_\delta,H)$ & $e^{-t\tilde{\mathcal{L}}(D_\delta,H)}$ & {T}ransition-probability matrix for a \\
& & Markov chain on an absorption-scaled graph \\ \hline
$P_l(D_\delta, H)$ & $I-t\tilde{\mathcal{L}}(D_\delta,H)$ & {L}inearized transition-probability matrix for a \\
& & Markov chain on an absorption-scaled graph\\ \hline
$\tilde{A}$ & {$\begin{pmatrix} A & \vec{0} \\ \vec{\delta}^{\T}& 0 \end{pmatrix}$} & {A}djacency matrix of a graph with an absorbing state\\
\hline
$\tilde{P}$ & & {T}ransition-probability matrix that is induced by $\tilde{A}$ \\ \hline
{$Q$} & {$A(W + D_\delta)^{-1}$} & {Submatrix of the transition-probability matrix $\Tilde{P}$} \\
& & {for transitions between non-absorbing states} \\
\hline 
$N$ & $(I - Q)^{-1}$ & Fundamental matrix of a discrete-time\\ 
& &  absorbing Markov chain \\ \hline
$\vec{t}$ & $N^T \vec{1}$ & {T}he entry $t_i$ of $\vec{t}$ is the expected number of transitions before \\
& & absorption when the initial state is $i$ \\ \hline
$\hat{N}$ & $ND_{\vec{t}}^{-1}$ & {N}ormalized fundamental matrix of a discrete-time\\
& &  absorbing {Markov chain} \\ \hline
$P_\delta$ & $P_l (D_\delta,I,1)$ & {S}pecial case of $P_l(D_\delta,H)$ with $H = I$ and $t = 1$\\ \hline
$\pi_\delta^{(a)}$ & $\hat{N}\pi_0$ & {T}he quantity $(\pi_\delta^{(a)})_i$ is the probability that node $i$ is the last node \\
 & & before absorption if the initial distribution is $\pi_0$  \\ \hline
$L^{(a)}(M,A,\vec{\delta},\pi_0)$ &  $L(M,P_\delta,\pi_\delta^{(a)})$ & {M}ap function for a Markov chain with an absorbing state\\ \hline \hline
\end{tabular}
\end{centering}
\caption{Summary of our notation.}
\label{table:notation}
\end{table}

\subsection{The map function and the standard InfoMap algorithm}\label{sec:Map}

In this subsection, we define an objective function, which is called the ``map function'', that the InfoMap algorithm seeks to minimize \cite{bohlin2014community, rosvall2009map,rosvall2008maps,infomap2023}. 

\begin{definition}{\bf (Map function)}\label{def:Map1}
Let $P = (p_{ij})_{i,j \in \{1,\ldots,n\}}$ be a stochastic matrix, and let $\pi = (\pi_1,\ldots,\pi_n)^{\T}$ be a probability distribution. For a partition $M = \{M_1,\ldots,M_m\}$ of the set of nodes of a network, we define the probability of a transition out of community $M_i$ with an initial state from the distribution $\pi$ by $q_{i\curvearrowright} := \sum_{j \in M_i, k \notin M_i} \pi_j p_{kj}$ and the probability of a transition into community $M_i$ with an initial state from the distribution $\pi$ by $q_{i\curvearrowleft} := \sum_{k \in M_i, j \notin M_i} \pi_j p_{kj}$. {The} probability of a transition into a community {that is different from the} community of an initial state from the distribution $\pi$ {is} $q_\curvearrowleft := \sum_{i \in \{1,\ldots,m\}} q_{i\curvearrowleft}$. The \emph{optimal mean encoding length that is associated with {the probabilities of transitions into a different community}} is the entropy\footnote{The entropy of a distribution with strictly positive probabilities $p_1, \ldots,p_r$ is $\mathcal{H}(\{p_1,\ldots, p_r\}) := -\sum_i p_i\log_2(p_i)$.} $\mathcal{H}(\mathcal{Q})$, where the distribution $\mathcal{Q}$ is
\begin{equation*} 	
	\mathcal{Q} := \left(q_{1\curvearrowleft}/q_\curvearrowleft, \ldots, q_{m\curvearrowleft}/q_\curvearrowleft \right)^{\T} \,.
\end{equation*}	
The \emph{optimal mean encoding length that is associated with the probabilities of transitions {out of community $M_i$}} is the entropy $\mathcal{H}(\mathcal{P}^i)$, where the probability distribution $\mathcal{P}^i$ is 
\begin{equation*}
	\mathcal{P}^i := { \left( q_{i \curvearrowright}/p_\circlearrowright^i, \pi_{k_1}/p_\circlearrowright^i, \ldots, \pi_{k_{{n_i}}}/p_\circlearrowright^i  \right)
	^{\T}\,, }
\end{equation*}	
where $M_i = \{k_1,\ldots, k_{{n_i}}\}$ (with $k_1 < \cdots < k_{{n_i}}$) {and $p_\circlearrowright^i := q_{i\curvearrowright } + \sum_{j\in M_i} \pi_j$ is a normalization factor.\footnote{{If $\pi$ is the stationary distribution corresponding to $P$, then $\pi_{k_1},\ldots,\pi_{k_{{n_i}}}$ are the associated probabilities that a transition ends at a node in community $M_i$ and we can interpret the probabilities in $\mathcal{P}^{i}$ as the probabilities of transitions that start in community $M_i$ and either leave it or remain in community $M_i$.}}} For the partition $M$, the \emph{map function that is associated with $P$ and $\pi$} is 
\begin{equation}\label{eqn:mapeqn}
	L(M,P,\pi) :=  q_\curvearrowleft \mathcal{H}(\mathcal{Q}) + \sum_{i=1}^m p_\circlearrowright ^i \mathcal{H}(\mathcal{P}^i)\, .
\end{equation}	
\end{definition}

\begin{definition}{\bf (Standard map function)}\label{def:Map11}
Let $P$ be a regular matrix (i.e., some power of $P$ has only positive entries), and let $\pi$ be its corresponding stationary distribution. We define $L(M,P)$ by $ L(M,P,\pi)$, and we refer to $L(M,P)$ as a \emph{standard map function}. 
\end{definition}

{The standard map function (see \cite{rosvall2008maps})} at a partition of a node set {is} based on entropies that are associated with codes that describe three types of transitions: (1) {transitions into a {different} community, (2) transitions {that end in a given community, and (3) transitions that start in a given community and then leave it. The quantity $\mathcal{H}(\mathcal{Q})$ in (\ref{def:Map1}) is the entropy of a distribution $\mathcal{Q}$ that is associated with transitions into a different community, and the $\mathcal{H}(\mathcal{P}^i)$ terms {in (\ref{def:Map1})} are entropies of the distributions $\mathcal{P}^i$ that are associated with transitions that either end in community $M_i$ or start in $M_i$ and then leave it. One can interpret these entropies as optimal mean encoding lengths (see Definition \ref{def:Map1}) in the sense of Shannon's source-coding theorem \cite{cover2012elements}. Shannon's source-coding theorem states that if $X$ is a random variable with finitely many states and $p$ is a probability mass function with entropy $\mathcal{H}(p)$, then the mean length of a code that describes the states of $X$ cannot be smaller than $\mathcal{H}(p)$. Additionally, as the size of the set of states of $X$ becomes infinite, one can approach the lower bound arbitrarily closely. Therefore, we refer to $\mathcal{H}(p)$ as the ``optimal mean encoding length'' (see Theorem 6 in \cite{shannon1948mathematical}), and we regard the map function as a weighted sum of optimal mean encoding lengths for one-step transitions between and within communities.

Let $A = (a_{ij})_{i,j\in\{1,\ldots,n\}}$ be the adjacency matrix of a directed and weighted graph with node set $\{1,\ldots,n\}$. The adjacency-matrix element $a_{ij}$ encodes the weight of the edge from node $j$ to node $i$, and $a_{ij} = 0$ implies that there is no edge from $j$ to $i$. The map function $L(M)$ in the following definition corresponds to the case in which the Markov chain that is induced by $A$ is regular.\footnote{A Markov chain is \emph{regular} if its associated transition-probability matrix is regular.} 

\begin{definition}{\bf (Map function associated with an adjacency matrix $A$)}\label{def:Map2}
Let $A$ be an adjacency matrix such that $AW^{-1}$ is regular, where $W:= \diag\{\omega_1,\ldots,\omega_n\}$ and $\omega_j := \sum_i a_{ij} \neq 0$ for $j \in \{1,\ldots,n\}$ is the out-degree of node $j$. The \emph{map function $L(M)$ that is associated with $A$} is
\begin{equation*}
	L(M) := L(M,AW^{-1}) \, .
\end{equation*}	
\end{definition}

The map function $L(M)$ measures the strength of the community structure of a partition $M = \{M_1,\ldots, M_m\}$ of the set of nodes. Intuitively, if the term $q_\curvearrowleft \mathcal{H}(\mathcal{Q})$ in (\ref{eqn:mapeqn}) is small, then the connections between communities are sparse. If the terms $p_\circlearrowright ^i \mathcal{H}(\mathcal{P}^i)$ are small, then there are dense intra-community connections \cite{rosvall2008maps, schaub2012encoding}. Therefore, we expect that minimizing $L(M)$ yields a partition with dense connections within communities and sparse connections between communities. The {standard InfoMap} algorithm attempts to minimize $L(M)$ using a greedy approach \cite{edler2017mapping} that is reminiscent of the Leiden algorithm \cite{traag2019louvain}.

\subsection{Markov time sweeping} \label{sec:Field}
 
In this subsection, we present Markov time sweeping for a version of the map function that includes a resolution parameter for community detection \cite{schaub2012encoding}. This resolution parameter, which amounts to a 
{``Markov time''} (which we will explain shortly), allows us to tune the sizes of the communities that we obtain using InfoMap. 

The map function \eqref{eqn:mapeqn} is associated with one-step transitions of a random walk. Schaub et al.~\cite{schaub2012encoding} incorporated Markov time sweeping\footnote{Markov sweeping time had been used previously in contexts other than InfoMap \cite{delvenne2010stability,lambiotte2008laplacian}.} into InfoMap to tune the time scales of transitions by encoding transitions with steps of any length $t > 0$. We think of the Markov chain that is determined by $T_J := AW^{-1}$ (with $AW^{-1}$ specified as in Definition \ref{def:Map2}) as having time steps of size $1$. Markov time sweeping uses the transition-probability matrix of a continuous-time Markov chain in which the time step is $t$ instead of $1$. We refer to the time $t$ as a ``Markov time''. Markov times $t < 1$ yield an encoding by the map function at a smaller transition time than $t = 1$. This, in turn, yields small communities. By contrast, Markov times $t > 1$ yield an encoding at a larger transition time than $t = 1$, so the encoding is able to capture transitions of a random walker that take more than one step. We thereby obtain large communities. See \cite{kheirkhahzadeh2016efficient} for further discussion of encodings and Markov times. 

As discussed in \cite{kheirkhahzadeh2016efficient}, for $t < 1$, one can also consider the linearization
\begin{equation}\label{eqn:linearization}
	e^{-t(I - T_J)} \approx I - t(I - T_J)= (1 - t)I + t T_J
\end{equation}
as an input of InfoMap. The matrix $(1-t)I + t T_J$ has diagonal elements that are all equal to $1 - t$ if we assume that $a_{ii} = 0$ for $i \in \{1,\ldots,n\}$. These diagonal elements correspond to self-edges with weight $1 - t$. If we use the standard map function with the transition-probability matrix $(1 - t)I + t T_J$ instead of $T_J$, we recover the map function in Definition \ref{def:Map2} by setting $t = 1$.

\section{Markov time sweeping and adaptations of InfoMap to absorbing random walks} \label{sec:scaledGrahExtension}

In this section, we present the adaptations of InfoMap that we use to explore how absorption in a dynamical process can affect community structure. To illustrate the implicit impact of absorption on community structure, consider an absorbing random walk on an undirected and unweighted line network with four nodes (see Figure \ref{fig:path_intro}).
We represent an absorbing state of an absorbing random walk on a graph as a {node\footnote{Throughout our paper, our figures do not show the nodes for the absorbing states that are associated with absorbing graphs. Our figures only show nodes that are associated with transient {(i.e., non-absorbing)} states of the corresponding absorbing random walks.}} with out-degree $0${. For} a given node, we refer to the sum of the weights of the edges between the node and absorbing states as the ``absorption rate'' of the node. Suppose that the absorption rate 
of node 2 in Figure \ref{fig:path_intro} is much larger than the absorption rates of the other nodes. The large absorption rate of node 2 is a barrier to the absorbing random walk, as transitions from nodes in the set $\{3,4\}$ to nodes in the set $\{1\}$ (and vice versa) are unlikely. The local dynamics (specifically, the absorption at the nodes) induces a partition of the node set that we can interpret as an {\it effective community structure} of the network. Such effective community structure can be rather different from the community structure that one detects based on network structure alone.


\begin{figure}[H]
    \centering
    \includegraphics[scale=0.6]{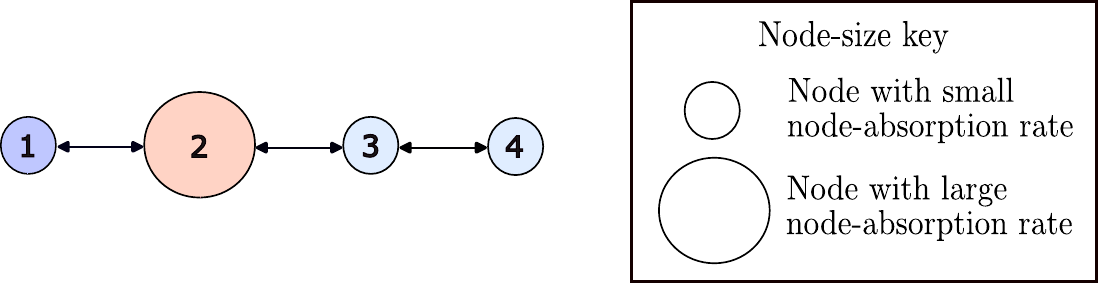}
    \caption{Consider an absorbing random walk on the depicted four-node network, and suppose that the absorption rate of node 2 is much larger than the absorption rates of the other nodes. Detecting communities via modularity maximization or the standard InfoMap algorithm produces a partition of the network into a single community that includes all nodes. However, the flow of an absorbing random walk is trapped in either the set $\{1\}$ (in dark blue) or the set $\{3,4\}$ (in light blue). Consequently, a partition that separates node 1 from nodes 3 and 4 better captures the dynamics of an absorbing random walk than a partition of the network into a single community.}
    \label{fig:path_intro}
\end{figure}

We now introduce adaptations of InfoMap that account for the absorption rates of the nodes of a network. Our approach uses {\it absorption-scaled graphs}, which arise naturally in the context of absorbing random walks \cite{jacobsen2018generalized}. 

\begin{definition}{\bf (Absorption-scaled graph)} \label{def:absorption_scaled}
Let $G$ be a directed and weighted graph with adjacency matrix $A = (a_{ij})_{i,j \in \{1,\ldots,n\} }$, where $a_{ij}$ encodes the weight of the edge from node $j$ to node $i$. Let $\vec{d} = (d_1,\ldots,d_n)^{\T}$ be a vector (which we call an ``absorption-rate vector'') with positive entries that we call the ``absorption rates''. We define the \emph{absorption-scaled graph that is associated with the pair} $(G,\vec{d})$ as the graph $\tilde{G}$ with adjacency matrix $\tilde{A} := AD^{-1}$, where $D := \diag\{d_1,\ldots,d_n\}$. 
\end{definition}


\begin{figure}[H]
    \centering
    \subfloat[][$(G,\vec{d})$]{\includegraphics[scale=0.5]{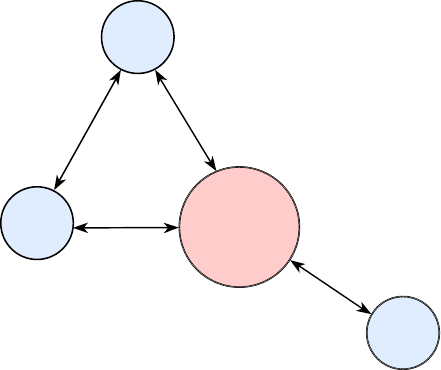}}
    \quad \quad \quad
    \subfloat[][$\tilde{G}$]{\includegraphics[scale=0.5]{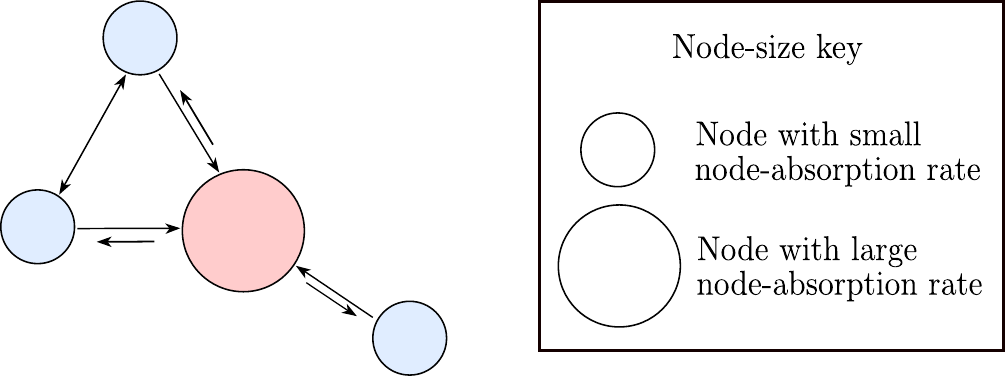}}
    \caption{An absorption-scaled graph. (a) A graph $G$ with absorption-rate vector $\vec{d}$. The pink node has a large absorption rate. (b) The associated absorption-scaled graph $\tilde{G}$, where the arrow length is proportional to the corresponding edge weight. 
    }
    \label{fig:absorption_graph}
\end{figure}

We now define some important mathematical objects that we use in our adaptations of InfoMap. 

\begin{definition}\label{def:abs_nodes}
Let  $A = (a_{ij})_{i,j \in \{1,\ldots,n\} }$ be the adjacency matrix of a graph $G$. Let $\vec{\delta} = (\delta_1, \dots, \delta_n)^{\T}$ be a vector with positive entries, where $\delta_i$ is the \emph{node-absorption rate} of node $i$ and is specified independently from the matrix $A$. We refer to the vector $\vec{\delta}$ as the \emph{node-absorption-rate vector}. Let $D_\delta := \diag\{\vec{\delta}\}$ and consider a diagonal matrix $H = \diag\{h_1, \ldots, h_n\}$, where $h_i \geq 0$. The \emph{scaled rate vector} $\vec{d}_{\mathrm{s}}(D_\delta, H)$ is given by the diagonal of $HW + D_\delta$.
\end{definition}

The node-absorption-rate vector $\vec{\delta}$ in Definition \ref{def:abs_nodes} does not depend on the adjacency matrix $A$, whereas the scaled rate vector $\vec{d}_{\mathrm{s}}(D_\delta,H)$ depends on the out-degrees $\omega_i$ through $H$. In particular, $\vec{\delta} = \vec{d}_{\mathrm{s}}(D_\delta, \mathbf{0})$ when $H = \mathbf{0}$. We use the notation $\vec{d}$ for a generic absorption-rate vector; it is not necessarily a node-absorption-rate vector or a scaled rate vector.

Our adaptations of InfoMap use Markov time sweeping (see Section \ref{sec:Field}) on absorption-scaled graphs. Consider a directed and weighted graph $G$ with adjacency matrix $A$ and node-absorption-rate vector $\vec{\delta} = (\delta_1, \dots, \delta_n)^{\T}$. We consider a family of absorption-scaled graphs with absorption-rate vectors that are equal to scaled rate vectors $\vec{d}_{\mathrm{s}}(D_\delta,H)$, where $H = \diag\{h_1,\ldots,h_n\}$ and $h_i\geq 0 $. 

Let $\tilde{G}(D_\delta, H)$ denote the absorption-scaled graph that is associated with the pair $(G,\vec{d}_{\mathrm{s}}(D_\delta,H))$. The unnormalized graph Laplacian matrix of $\tilde{G}(D_\delta, H)$ is 
\begin{equation}\label{eqn:LDH}
	\tilde{\mathcal{L}}(D_\delta,H) := (W-A)(HW+D_\delta)^{-1} \, .
\end{equation}
For any Markov time $t > 0$, the {transition-probability matrix that is associated with the infinitesimal generator} $-\tilde{\mathcal{L}}(D_\delta,H)$ is 

\begin{equation}\label{eqn:PDHtexp}
	 P_{e}(D_\delta,H,t) := e^{-t \tilde{\mathcal{L}}(D_\delta,H)} \, .    
\end{equation}
For a Markov time $t$, the {linearization of} $P_{e}(D_\delta,H,t)$ is 
\begin{equation}\label{eqn:PDHt}
	 P_{l}(D_\delta,H,t) := I - t\tilde{\mathcal{L}}(D_\delta,H) = \left(I - t W(HW+D_\delta)^{-1}\right) + tA(HW+D_\delta)^{-1} \, .    
\end{equation}
For this linearization, we require that $0<t<1/\max_i\{\omega_i/(h_i\omega_i + \delta_i)\}$ to ensure that $P_l(D_\delta,H,t)$ is a transition-probability matrix.  

We use the matrix $H$ in Definition \ref{def:abs_nodes} because this matrix allows us to tune the relative effects of the edge weights and the node-absorption rates on the communities that we detect using our adaptations of InfoMap. In Algorithms \ref{alg:info_abs_partA} and \ref{alg:info_abs_partB}, we summarize our adaptations of InfoMap. 


\begin{subalgorithms}

\begin{algorithm}
\caption{InfoMap for absorbing random walks with a linear input.}
\label{alg:info_abs_partA}
\begin{algorithmic}[1]
 \renewcommand{\algorithmicrequire}{Input:}
 \renewcommand{\algorithmicensure}{Output:}
 \REQUIRE An adjacency matrix $A=(a_{ij})_{i,j\in \{1,\ldots,n\} }$ of a directed and weighted graph, a node-absorption-rate vector $\vec{\delta} = (\delta_1,\ldots,\delta_n)^{\T}$ with strictly positive entries, and a diagonal matrix $H$ with non-negative entries. \medskip
 \ENSURE A partition $M$ of the set of nodes that minimizes $L(M,P_l(D_\delta,H,t))$ for a Markov time $t$.
\medskip \medskip
	\STATE Construct the unnormalized graph Laplacian matrix $\tilde{\mathcal{L}}(D_\delta,H) = (W-A)(HW + D_\delta)^{-1}$ for the absorption-scaled graph $\tilde{G}(D_\delta,H)$. \medskip
	\STATE Choose a Markov time such that
	\begin{equation}\label{eqn:feasible_alg}
	0<t<1/\max_i\{\omega_i/(h_i\omega_i + \delta_i)\} \,.
	\end{equation} \medskip
	\STATE {Apply the standard InfoMap algorithm to minimize  $L\left({M}, P_l(D_\delta,H,t) \right)$.} 
\end{algorithmic}
\end{algorithm}

\begin{algorithm}
\caption{InfoMap for absorbing random walks with an exponential input.}
\label{alg:info_abs_partB}
\begin{algorithmic}[1]
 \renewcommand{\algorithmicrequire}{Input:}
 \renewcommand{\algorithmicensure}{Output:}
 \REQUIRE An adjacency matrix $A = (a_{ij})_{i,j\in \{1,\ldots,n\} }$ of a directed and weighted graph, a node-absorption-rate vector $\vec{\delta} = (\delta_1,\ldots,\delta_n)^{\T}$ with positive entries, and a diagonal matrix $H$ with non-negative entries. \medskip
 \ENSURE A partition $M$ of the set of nodes that minimizes $L(M,P_e(D_\delta,H,t))$ for a Markov time $t$.
\medskip \medskip
	\STATE Construct the unnormalized graph Laplacian matrix $\tilde{\mathcal{L}}(D_\delta,H) = (W-A)(HW + D_\delta)^{-1}$ for the absorption-scaled graph $\tilde{G}(D_\delta,H)$. \medskip
	\STATE Choose any Markov time $t > 0$.  \medskip
	\STATE {Apply the standard InfoMap algorithm to minimize  $L\left({M}, P_{{e}}(D_\delta,H,t) \right)$.}  
\end{algorithmic}
\end{algorithm}

\end{subalgorithms}

In our adaptations of InfoMap to absorbing random walks, we introduce a family of associated absorption-scaled graphs and then apply Markov time sweeping to these absorption-scaled graphs. To illustrate how the node-absorption rates impact the communities that we detect, consider the matrix $P_{{l}}$ in Algorithm \ref{alg:info_abs_partA}. In the expression for $P_{{l}}(D_\delta,H,t)$ in \eqref{eqn:PDHt}, the term
\begin{equation*} 
	I - t W(HW + D_\delta)^{-1} = \diag\left\{1 - \frac{t\omega_i}{h_i\omega_i + \delta_i} \right\}
\end{equation*}	
creates self-edges that are positively correlated with the node-absorption rates $\delta_i$. This correlation reflects the idea that a random walk gets stuck longer in nodes with larger node-absorption rates.

If we let $D_\delta = \delta_* I$ and $\omega_i \geq 1$ (with $i \in \{1,\ldots,n\}$) and set $h_i = \delta_*(\omega_i - 1)/\omega_i$ and $t = \delta_*$, then $P_{{l}}(\delta_* I, H, \delta_*) = AW^{-1}$. We thereby recover the input of the standard InfoMap algorithm when all of the absorption rates are the same. By letting $H = hI$ and $t = h$, we obtain 
\begin{equation}
    \lim_{\norm{\vec{\delta}} \rightarrow 0} P_{{l}}(D_\delta,hI,h) = \lim_{\norm{\vec{\delta}} \rightarrow 0}(I - h W(hW + D_\delta)^{-1}) + hA(hW + D_\delta)^{-1} = AW^{-1} \, , 
\end{equation}
which again recovers the input of the standard InfoMap algorithm.

\section{A map function $L^{(a)}$ for Markov chains with an absorbing state} \label{sec:redefMap}

\subsection{A map function for an absorbing {random walk}}\label{sec:La}

It is natural to ask if one can interpret map functions that are associated with our adaptations of InfoMap in terms of corresponding Markov chains with an absorbing state. {We {take} a step towards answering this question by defining a map function $L^{(a)}$ for Markov chains with an absorbing state. The main results of this subsection are that (1) the map function $L^{(a)}$ is the map function that is associated with {our adaptations of InfoMap in Algorithm \ref{alg:info_abs_partA} with $H = I$} and that (2) the map function $L^{(a)}$ converges to the standard map function as the absorption rates approach $0$.}

\subsubsection{Construction of a map function for an absorbing random walk}
\label{sec:pdelta_pa}

We assume that the Markov chain that is associated with the adjacency matrix $A$ is regular, and we add an absorbing state (i.e., a new node with out-degree $0$) and node-absorption rates $\delta_1,\ldots, \delta_n$, which are the transition rates from states that are associated with $A$ to the absorbing state. The adjacency matrix of the absorbing Markov chain is 
\begin{equation*}
	\Tilde{A} = \begin{pmatrix} A & \vec{0} \\ \vec{\delta}^{\T}& 0 \end{pmatrix}\,,
\end{equation*} 
where $\vec{\delta} = (\delta_1,\ldots,\delta_n)^{\T}$. From $\Tilde{A}$, we obtain the transition-probability matrix
\begin{equation}\label{eqn:PtildeAbs}
	\tilde{P} = \begin{pmatrix} Q & \vec{0} \\ \vec{r}^{\T} & 1  \end{pmatrix}\,,
\end{equation}
{where the transition probabilities to absorption are the entries of $\vec{r} = (\delta_1/(\omega_1 + \delta_1),\ldots, \delta_n/(\omega_n + \delta_n))^{\T}$, with $\omega_j = \sum_i a_{ij}$ (for $j \in \{1,\ldots,n\}$),} and $Q = A(W + D_\delta)^{-1}$, with $D_\delta = \diag\{\vec{\delta}\}$ and $W = \diag\{\omega_1,\ldots,\omega_n\}$. The Markov chain with the transition-probability matrix $\tilde{P}$ is the absorbing Markov chain that is associated with $A$ and $\delta_1,\ldots, \delta_n$.

By Definition \ref{def:Map1}, we know that if $P$ is a regular transition-probability matrix and $\pi$ is a probability mass function, then $L(M,P,\pi)$ depends only on $M$, $P$, and $\pi$. We write $L(M,P,\pi)$ as $L(M,P)$ if $\pi$ is the unique stationary distribution of $P$. The quantity $L(M,P,\pi)$ is a weighted sum of optimal mean encoding lengths for one-step transitions (with probabilities in $P$ and starting from the distribution $\pi$) between and within communities. We define a map function $L^{(a)}$ for the absorbing random walk that is associated with \eqref{eqn:PtildeAbs} by $L(M,P_\delta,\pi_\delta^{(a)})$ for an appropriate distribution $\pi_\delta^{(a)}$ (see Section \ref{sec:distr_pdelta}) and an appropriate transition-probability matrix $P_\delta$ (see Section \ref{subsec:P_delta}).

\paragraph{The distribution $\pi_\delta^{(a)}$}\label{sec:distr_pdelta}

We define a distribution $\pi_\delta^{(a)}$ {with} the desirable property of recovering the stationary distribution $\pi$ of a Markov chain without {absorption} when the absorption rates approach $0$. 
{(This Markov chain has {the} transition-probability matrix $AW^{-1}$.)} That is, 
\begin{equation}\label{eqn:limpidelta}
	\lim_{\norm{\vec{\delta}} \rightarrow 0} \pi_\delta^{(a)} = \pi \,.
\end{equation}

{Let $N = (I - Q)^{-1} = \sum_{k = 0}^\infty Q^k$. The entry $n_{ij}$ of $N$ gives {the} expected number of visits to {node $i$ that start} from node $j$. {The matrix $N$ is the} \emph{fundamental matrix} of the corresponding absorbing Markov chain \cite{kemeny1983finite}. In particular, $N$ is the fundamental matrix of the absorbing Markov chain with transition-probability matrix $\tilde{P}$.} Consider the normalized fundamental matrix 
\begin{equation} \label{eqn:Nhat}
	\hat{N} = ND_{\vec{t}}^{-1} \,,
\end{equation}
where $\vec{t} = N^{\T}\vec{1} = (t_1,\ldots, t_n)^{\T}$, with $\vec{1} := (1,\ldots,1)^{\T}$, is the vector whose entries give the expected numbers of steps before absorption starting from each non-absorbing state. For each node pair $(i,j)$, the entry $n_{ij}/t_j$ of $\hat{N}$ gives the probability that node $i$ is the last node before absorption if we start at node $j$. This probability depends on the node-absorption rates. Given an initial distribution $\pi_0$, we obtain the distribution 
\begin{equation} \label{eqn:pideltaAbs}
	\pi_\delta^{(a)} = \hat{N}\pi_0 \, .    
\end{equation}

The following proposition states that equation \eqref{eqn:limpidelta} holds. 

\begin{proposition}\label{prop:pi_abs}
Suppose that the Markov chain with transition-probability matrix $AW^{-1}$ is regular. Let $\vec{\delta}$ be a node-absorption-rate vector in which all entries are strictly positive, and let $D_\delta := \diag\{\vec{\delta}\}$. Let $N = (I - A(W+D_\delta)^{-1})^{-1}$ be the fundamental matrix of the absorbing Markov chain that is associated with $A$ and $\delta_1,\ldots,\delta_n$. Additionally, let $D_{\vec{t}}$ be the diagonal matrix with the column sums of $N$ in its diagonal, and let $\pi$ be the stationary-distribution vector that is associated with $AW^{-1}$. 

It then follows that
\begin{equation*}
	\lim_{\norm{\vec{\delta}} \rightarrow 0} ND_{\vec{t}^{-1}} = \pi \vec{1}^{\T}\, . 
\end{equation*}	

\end{proposition}

\begin{proof}
Consider the vector 1-norm $\norm{\vec{b}}_1 = \sum_i |b_i|$ and its induced matrix norm $\norm{B}_1 = \max_j \sum_i |b_{ij}|$. Fix $\epsilon > 0$. Because $\lim_{n \rightarrow \infty} (AW^{-1})^n = \pi \vec{1}^{\T}$, it follows that there is a positive integer $N_1$ such that
\begin{align}
	\norm{(AW^{-1})^n-\pi \vec{1}^{\T}}_1 < \epsilon  \quad \textnormal{ for }\quad  n\geq N_1 \,.   
\end{align}
In particular, 
\begin{equation}\label{eqn:aw_Lambda}
	(AW^{-1})^{N_1} = \pi \vec{1}^{\T} + \Lambda\,,
\end{equation}
where $\norm{\Lambda}_1 <\epsilon$. Let $Q := A(W+D_\delta)^{-1}.$ 
Because $\norm{Q^j}_1 \leq \norm{Q}_1^j \leq \norm{AW^{-1}}_1^j = 1$ and $t_i \rightarrow \infty$ as $\norm{\vec{\delta}}_1 \rightarrow 0$, it follows that 
\begin{equation}
    \lim_{{\norm{\vec{\delta}}_1} \rightarrow 0} \norm{\sum_{j<N_1} Q^j D_{\vec{t}}^{-1} }_1 = 0 \,.
\end{equation}
Additionally, $\lim_{\norm{\vec{\delta}}_1 \rightarrow 0} Q^{N_1} = (AW^{-1})^{N_1}$. Therefore, there is a $\eta_0 > 0$ such that 
\begin{align}\label{eqn:lesser_N1}
	\left\|{\sum_{j<N_1} Q^j D_{\vec{t}}^{-1} }\right\|_1 <\epsilon
\quad \textnormal{ and }\quad Q^{N_1} = (AW^{-1})^{N_1} + \Delta
\end{align}
whenever $0<\norm{\vec{\delta}}_1 <\eta_0$, where $\norm{\Delta}_1 < \epsilon$. Let $\Delta' := \Lambda + \Delta$. Using (\ref{eqn:aw_Lambda}) and (\ref{eqn:lesser_N1}), we obtain
 \begin{align}\label{eqn:greater_N1}
	 \sum_{j\geq N_1} Q^j D_{\vec{t}}^{-1} & = Q^{N_1}\sum_{j\geq 0} Q^j D_{\vec{t}}^{-1} \notag \\  &= Q^{N_1}  ND_{\vec{t}}^{-1} \notag \\  &= (\pi \vec{1}^{\T} + \Delta')ND_{\vec{t}}^{-1} \notag \\  &= \pi \vec{1}^{\T} ND_{\vec{t}}^{-1} + \Delta' ND_{\vec{t}}^{-1} \notag \\  &= \pi\vec{t}^{\T}D_{\vec{t}}^{-1} 
	 + \Delta' ND_{\vec{t}}^{-1}\notag \\  &= \pi \vec{1}^{\T} + \Delta' ND_{\vec{t}}^{-1}\,,
\end{align} 
where $\norm{\Delta' ND_{\vec{t}}^{-1} }_1 \leq \norm{\Delta'}_1\norm{ ND_{\vec{t}}^{-1}}_1 < 2\epsilon$.
 From (\ref{eqn:lesser_N1}) and (\ref{eqn:greater_N1}), it follows for $0 < \norm{\vec{\delta}}_1 < \eta_0$ that 
 \begin{equation*}
 	\norm{ND_{\vec{t}}^{-1}-\pi\vec{1}^{\T}}_1 \leq \norm{\sum_{j<N_1} Q^j D_{\vec{t}}^{-1}}_1 +\norm{\Delta' ND_{\vec{t}}^{-1}}_1 < 3\epsilon\,.
\end{equation*}	
\end{proof}

\paragraph{The transition-probability matrix $P_\delta$} \label{subsec:P_delta}

Let $P_\delta$ denote the linearization (\ref{eqn:PDHt}) with $H = I$ and $t = 1$. That is, 
\begin{equation}\label{eqn:PdeltaRedef}
	P_\delta :=  P_l(D_\delta, I, 1) = (I - W(W+D_\delta)^{-1}) + A(W+D_\delta)^{-1} = D_{\vec{r}} + Q \, ,  
\end{equation}
where $D_{\vec{r}} = \diag\{\vec{r}\}$ with $\vec{r}$ as in \eqref{eqn:PtildeAbs}. Our choice of $P_\delta$ is motivated by the property
\begin{align}\label{eqn:limPdelta}
   \lim_{\norm{\vec{\delta}} \rightarrow 0} P_\delta = P =  AW^{-1} \, .
\end{align}
From (\ref{eqn:limPdelta}), we recover the transition-probability matrix of the regular Markov chain that is induced by $A$ in the limit in which there is no absorption. {The diagonal entries of $P_{\delta}$ correspond to the one-step absorption probabilities $\delta_i/(\omega_i + \delta_i)$ in the absorbing Markov chain that is associated with the transition-probability matrix $\tilde{P}$ (see \eqref{eqn:PtildeAbs}). The following} proposition states that the time to self-transitions (i.e., a transition from a node to itself) of the Markov chain that is associated with $P_\delta$ is equal to the time to absorption in the absorbing Markov chain that is associated with $\tilde{P}$.

\begin{proposition}\label{prop:absWait}
Let $\{X_n\}_{n\in\mathbb{N}}$ be the Markov chain with transition-probability matrix $P_\delta = P_l(D_\delta,I,1) = (p_{ij}^{(1)})_{i,j\in\{1,\ldots,n\}}$. Define the random variable 
\begin{equation}
	T_j := \min\{n: X_n = X_{n-1} \text{ and } X_0 = j\}\,.
\end{equation}
Let $\theta_j = \mathbb{E}(T_j)$ be the expected time to the first self-transition, and let $\vec{\theta} = (\theta_1,\ldots,\theta_n)^{\T}$. If $N = (I-Q)^{-1}$ is the fundamental matrix of the absorbing Markov chain that is associated with $A$ and $\delta_1,\ldots,\delta_n$, it follows that
\begin{equation}
	\vec{\theta}^{\T} = \vec{1}^{\T}N \, .
\end{equation}	
\end{proposition}

\begin{proof}
From the law of total expectation,
\begin{equation}\label{eqn:first_step}
	 \theta_j = \sum_{i\neq j} (\mathbb{E}(T_i)+1)p_{ij}^{(1)}+ p_{jj}^{(1)}  = \sum_{i\neq j} \theta_ip_{ij}^{(1)}+1\,. 
\end{equation}

Let $(P_\delta)_{\dg}$ denote the diagonal matrix with the same diagonal as $P_\delta$. We write (\ref{eqn:first_step}) as 
\begin{equation}\label{eqn:first_step2}
	 \vec{\theta}^{\T} [I - (P_\delta - (P_\delta)_{\dg}) ] = \vec{1}^{\T} \,.
\end{equation}
Because $P_\delta = D_{\vec{r}} + Q $, it follows from (\ref{eqn:first_step2}) that $Q = P_\delta - (P_\delta)_{\dg}$ and $\vec{\theta}^{\T} = \vec{1}^{\T}N$. 
\end{proof}

We now define a map function for the absorbing random walk that is associated with $A$ and $\delta_1,\ldots,\delta_n$.

\begin{definition}{{\bf (Map function for an absorbing random walk)
}}\label{def:La}
Let $A$ be the adjacency matrix of an absorbing random walk, and let $\vec{\delta}$ be the walk's node-absorption-rate vector. Let $M$ be a partition of the node set that is associated with $A$, and let $\pi_0$ be an initial probability distribution. We define the \emph{map function} $L^{(a)}(M,A,\vec{\delta},\pi_0) := L(M,P_\delta,\hat{N}\pi_0)$ {with $P_\delta$ in \eqref{eqn:PdeltaRedef} and $\hat{N}$ by \eqref{eqn:Nhat}.}
\end{definition}

As an instructive example, we calculate the map function $L^{(a)}$ for all possible partitions $M$ of the three-node network with node set  $\{1,2,3\}$ and adjacency matrix
\begin{equation}\label{eqn:A_3nodes}
	A =\begin{pmatrix} 0 & 1 & 1 \\ 0 & 0 & 0 \\ 1 & 1 & 0\end{pmatrix} \,.
\end{equation}
Intuitively, if the node-absorption rate of node $2$ is larger than the node-absorption rates of nodes $1$ and $3$, then node $2$ is in a different community than nodes $1$ and $3$ in the effective community structure. We want to check whether or not the partition with the minimum value of $L^{(a)}$ captures this intuition. 

Let $\vec{\delta} = (\delta_1,\delta_2,\delta_3)^{\T}$ be a node-absorption-rate vector. We fix $\delta_1 = \delta_3 = 0.1$ and vary $\delta_2$ in the interval $[0.1,10]$. In Figure \ref{fig:3nodesRedefMap}, we show the values of $L^{(a)}(M,A,\vec{\delta},\pi_0)$ for all five possible partitions $M$ of the set $\{1,2,3\}$ of nodes, where $A$ is specified in \eqref{eqn:A_3nodes} and $\pi_0$ is the uniform distribution on $\{1,2,3\}$. We always attain the minimum value of $L^{(a)}(M,A,\vec{\delta},\pi_0)$ for the partition $\{\{2\},\{1,3\}\}$, so we obtain this partition if we select the optimal encoding $L^{(a)}(M,A,\vec{\delta},\pi_0)$. Because $\delta_2 >\delta_1$ and $\delta_2 >\delta_3$, this result is consistent with the intuition that $\{\{2\},\{1,3\}\}$ is the effective community structure. 


\begin{figure}[H]
    \centering
    \includegraphics[scale= 0.5]{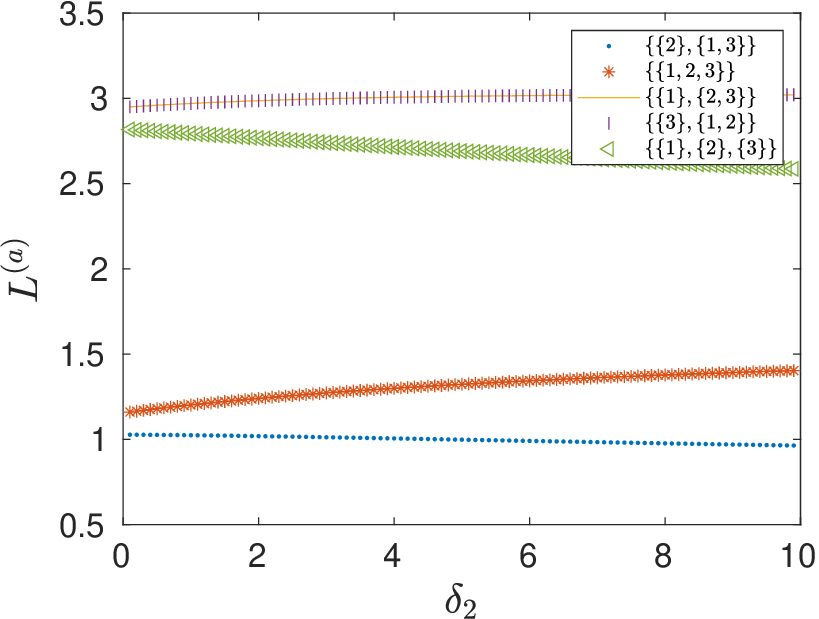}
    \caption{The {map function} $L^{(a)}(M,A,\vec{\delta},\pi_0)$ for all five possible partitions $M$ of 
    the three-node network with adjacency matrix (\ref{eqn:A_3nodes}). The node-absorption-rate vector is $\vec{\delta} = (\delta_1,\delta_2,\delta_3)^{\T}$, with $\delta_1 = \delta_3 = 0.1$ and $0.1 \leq \delta_2 \leq 10$. The initial distribution is $\pi_0 = (1/3,1/3,1/3)^{\T}$.
    }
    \label{fig:3nodesRedefMap}
\end{figure}

\subsubsection{Main results for the map function $L^{(a)}$} \label{sec:main_results_la}

Our first key result in the present subsubsection is that the standard map function $L(M,P_\delta)$, with $P_\delta$ as in \eqref{eqn:PdeltaRedef}, equals the map function $L^{(a)}(M,A,\vec{\delta},\pi_0)$ for an appropriate distribution $\pi_0$.

\begin{proposition} \label{prop:pa_ext}
Suppose that $P_\delta = P_l(D_\delta,I,1)= (I - W(W+D_\delta)^{-1}) + A(W + D_\delta)^{-1} = D_{\vec{r}} + Q$. Let $\pi_\delta^{(na)}$ be the stationary distribution of $P_\delta$, and let $\pi_0 := D_{\vec{t}}D_{\vec{r}}\pi_\delta^{(na)} = (\pi_{\delta,i}^{(na)} t_i\delta_i/(\omega_i+\delta_i))_i$. 
We have that
\begin{equation}
	L(M,P_\delta) = L^{(a)}(M,A,\vec{\delta},\pi_0) \, . 
\end{equation}	
\end{proposition}

\begin{proof}
Because $\pi_\delta^{(na)}$ is the stationary distribution of $P_\delta$, it follows that $(D_{\vec{r}} + Q)\pi_\delta^{(na)} = \pi_\delta^{(na)} $, which implies that
\begin{equation}
	\pi_\delta^{(na)} = \hat{N} D_{\vec{t}}D_{\vec{r}}\pi_\delta^{(na)} = \hat{N}\pi_0  \, .
\end{equation}
Because the entries of $\pi^{(na)}_\delta$ sum to $1$ and the columns of $\hat{N}$ each sum to $1$, we know that $\pi_0$ is a probability distribution.  Therefore, for this choice of $\pi_0$, it follows that 
\begin{equation}
	L(M,P_\delta) = L(M, P_\delta, \pi_\delta^{(na)}) = L(M, P_\delta, \hat{N}\pi_0) = L^{(a)}(M,A,\vec{\delta},\pi_0) \, ,
\end{equation}
where $L(M,P_\delta)$ is the standard map function with input $P_\delta$ and $ L^{(a)}(M,A,\vec{\delta},\pi_0)$ is the map function in Definition \ref{def:La}.
\end{proof}

{
\begin{remark*}
The ``na'' in the superscript of $\pi^{(na)}_{{\delta}}$ stands for ``non-absorbing''. (The probability distribution $\pi^{(na)}_{{\delta}}$ is the stationary distribution of the non-absorbing Markov chain with transition-probability matrix $P_\delta$.) 
\end{remark*}
}

Our second key result in the present subsubsection is that 
\begin{equation}\label{eqn:limLs}
	 L^{(a)}(M,A,\vec{\delta},\pi_0) \rightarrow L(M, P) \quad \text{as} \quad \norm{\vec{\delta}} \rightarrow 0
\end{equation}
for any $\pi_0$. This result follows from $\lim_{\norm{\vec{\delta}} \rightarrow 0} \pi_\delta^{(a)} = \pi$ (see Proposition \ref{prop:pi_abs}) and $\lim_{\norm{\delta} \rightarrow 0} P_\delta = P = AW^{-1}$. 
Therefore, the map function $ L^{(a)}(M,A,\vec{\delta},\pi_0)$ that is associated with the absorbing random walk converges to the map function $L(M,P)$ that is associated with the (non-absorbing) Markov chain in the limit $\norm{\vec{\delta}} \rightarrow 0$.

\subsection{Relating $\tilde{G}(D_\delta, I)$ to $\tilde{G}(D_\delta, \mathbf{0})$ } \label{sec:relating_G_delta}

{Our adaptation of InfoMap to absorbing random walks involves a family of absorption-scaled graphs $\tilde{G}(D_\delta, H)$, where $H$ is a scaling matrix that controls the relative importances of edge weights and node-absorption rates to community detection. The choice $H = \mathbf{0}$ corresponds to an absorption-scaled graph with an absorption-rate vector whose entries {are} the node-absorption rates. As discussed in \cite{jacobsen2018generalized}, {the absorption-scaled graph $\tilde{G}(D_\delta, H)$} with $H = \mathbf{0}$ is related to the fundamental matrix of the continuous-time absorbing random walk that is associated with $A$ and $\vec{\delta}$. This relationship arises through the absorption inverse, which is a particular generalized inverse of the {unnormalized} graph Laplacian matrix. Our results in Section \ref{sec:La} indicate that $H = I$ is also a distinguished choice.} It is natural to ask how the absorption-scaled graphs $\tilde{G}(D_\delta, \mathbf{0})$ and $\tilde{G}(D_\delta, I)$ are related. In this subsection, we find relationships between the Markov chains that are associated with the graph Laplacian matrices $\tilde{\mathcal{L}}(D_\delta,\mathbf{0})$ and $\tilde{\mathcal{L}}(D_\delta,I)$ through their associated fundamental matrices and absorption inverses. We describe these relationships, which are the main results of this subsection, in Propositions \ref{prop:z0_z1}, \ref{prop:L0L1}, and \ref{prop:d0_d1}. These results also yield connections between $\tilde{G}(D_\delta, \mathbf{0})$, $\tilde{G}(D_\delta, I)$, and the fundamental matrix $(\mathcal{L} + D_\delta)^{-1}$ through Propositions \ref{prop:group_inv} and \ref{prop:fund_and_absinv}.

We first look at the fundamental matrices of the discrete-time Markov chains that are associated with $\tilde{\mathcal{L}}(D_\delta, \mathbf{0})$, and $\tilde{\mathcal{L}}(D_\delta, I)$. Definition \ref{def:fm_reg} describes the fundamental matrix of a regular Markov chain. 

\begin{definition}{{\bf (Fundamental matrix)}}\label{def:fm_reg}
Let $P$ be a regular transition-probability matrix, and let $\vec{p}$ be its corresponding stationary distribution. The \emph{fundamental matrix $Z$ of the Markov chain that is associated with} $P$ is 
\begin{equation}
	Z = (I - P + \vec{p}\vec{1}^{\T})^{-1} \,. 
\end{equation}
\end{definition}

The entries of the matrix $Z - \vec{p}\vec{1}^T$ approximate the differences between the expected numbers of visits of the Markov chain that {is} associated with $P$ {and the expected numbers of visits of} the Markov chain that is associated with $\vec{p}\vec{1}^{\T}$. (See Section 4.3 of \cite{kemeny1983finite}.) 

The following proposition gives a relationship between the fundamental matrices of the discrete-time Markov chains that are associated with the graph Laplacian matrices $\tilde{\mathcal{L}}(D_\delta,\mathbf{0})$ and $\tilde{\mathcal{L}}(D_\delta, I)$. We prove this proposition in Appendix \ref{appendix}.

\begin{proposition}\label{prop:z0_z1}
Let $P_0 = AW^{-1}$ be the transition-probability matrix of the discrete-time Markov chain that is associated with $\tilde{\mathcal{L}}(D_\delta,\mathbf{0})$, and let $P_1 = (A+D_\delta)(W+D_\delta)^{-1}$ be the transition-probability matrix of the discrete-time Markov chain that is associated with $\tilde{\mathcal{L}}(D_\delta, I)$. Let $\pi$ and $\pi'$ be the stationary distributions that are associated with $P_0$ and $P_1$, respectively. Let $Z_i$ be the fundamental matrix that is associated with $P_i$ (with $i \in \{1,2\}$). Let $U := (1/(\vec{\delta}^{\T}\vec{u}))\vec{u}\vec{1}^{\T}$ and $\alpha := \vec{\delta}^{\T}\vec{u}/(\vec{w}^{\T}\vec{u}+\vec{\delta}^{\T}\vec{u})$, where $\vec{u} = (u_1,\ldots,u_n)^{\T}$ is a vector in the kernel $\Ker {(W - A)}$ with positive entries {$u_i$} such that $\sum_{i=1}^n u_i = 1$. We have that
\begin{equation}\label{eqn:pfZ0}
	Z_1 = W^{-1}(W+D_\delta)\left[Z_0 + \alpha(1-\alpha)\pi \vec{1}^{\T} - \alpha\left(Z_0D_\delta U + W\frac{\vec{u}\vec{\delta}^{\T}}{\vec{\delta}^{\T}\vec{u}}W^{-1}Z_0(I-\alpha D_\delta U)\right)\right] \, .
\end{equation}
\end{proposition}

\medskip

\begin{remark*}
The expression in {square brackets in} the right-hand side of (\ref{eqn:pfZ0}) 
is a fundamental matrix {that is associated with $P_0$~\cite{kemeny1983finite}.}
\end{remark*}

We now seek a connection between the unnormalized graph Laplacian matrices $\tilde{\mathcal{L}}(D_\delta, \mathbf{0})$ and $\tilde{\mathcal{L}}(D_\delta, I)$ through absorption inverses. Definition \ref{def:absinv} defines the absorption inverse of an unnormalized graph Laplacian matrix {and absorption-rate vector $\vec{d}$}. 

\begin{definition}{{\bf (Absorption inverse)}}\label{def:absinv}
Let $\tilde{\mathcal{L}}$ be the unnormalized graph Laplacian matrix of a strongly connected graph, and let $\vec{d}$ be an absorption-rate vector. Let $D:= \diag\{\vec{d}\}$, $N_{1,0} := \{ \vec{x} \in \mathbb{R}^n: D\vec{x}\in \range \tilde{\mathcal{L}} \}$, and $R_{1,0}:=\{D\vec{x}: \vec{x} \in \Ker \tilde{\mathcal{L}} \}$. An \emph{absorption inverse $\tilde{\mathcal{L}}^{\vec{d}}$ of $\tilde{\mathcal{L}}$ with respect to $\vec{d}$} is defined by the following properties:
\begin{align}
	\tilde{\mathcal{L}}^{\vec{d}}\tilde{\mathcal{L}}\vec{y} &= \vec{y} \quad \textnormal{ for } \quad  \vec{y} \in N_{1,0} \, ,    \notag \\
	\tilde{\mathcal{L}}^{\vec{d}}\vec{y} &= \vec{0} \quad \textnormal{ for }\quad  \vec{y} \in R_{1,0} \,.    
\end{align}
\end{definition}

The absorption inverse of a graph exists and is unique if the graph is strongly connected and the absorption rates are all positive. 
(See Theorem 2 in \cite{jacobsen2018generalized}.) 
{As described in \cite{jacobsen2018generalized}, the absorption inverse of a graph is closely related to the group inverse of an associated matrix.}

\begin{definition}{{\bf (Group inverse)}}\label{def:gp_inv}
Let $X$ be a {square} matrix such that $\rank(X) = \rank(X^2)$. The \emph{group inverse of} $X$ is the unique matrix $X^{\#}$ that satisfies 
\begin{align} \label{eqn:group_inv1}
	XX^{\#}X &= X \, ,    \notag \\
	X^{\#}XX^{\#} &= X^{\#} \, , \notag \\
	XX^{\#} &= X^{\#}X \, .
\end{align}
\end{definition}

\begin{proposition}[\cite{jacobsen2018generalized}, Proposition 2] \label{prop:group_inv}
Let $\tilde{\mathcal{L}}$ be the unnormalized graph Laplacian matrix of a strongly connected graph, let $\vec{d}$ be an absorption-rate vector, and let $D = \diag\{\vec{d}\}$. The following relationship holds:
\begin{equation*}
	\left(\tilde{\mathcal{L}}D^{-1}\right)^{\#} = D\tilde{\mathcal{L}}^{\vec{d}}  \, .
\end{equation*}	
\end{proposition}

{The absorption inverse $\tilde{\mathcal{L}}^{\vec{d}}$ is related both to the fundamental matrix $(\tilde{\mathcal{L}} + D)^{-1}$ of the continuous-time absorbing Markov chain that is associated with $\tilde{\mathcal{L}}$ and $d_1,\ldots,d_n$ and to the fundamental matrix of the associated discrete-time Markov chain.  Propositions \ref{prop:fund_and_absinv} and \ref{prop:compute_abs_inv} give two such relationships. See~\cite{jacobsen2018generalized} for the proofs of these propositions. We use these prior results to establish relationships between $\tilde{G}(D_\delta, \mathbf{0})$ and $\tilde{G}(D_\delta, I)$. 
We give these results in Propositions \ref{prop:L0L1} and \ref{prop:d0_d1}.}

\begin{proposition}[\cite{jacobsen2018generalized}, Proposition 4]\label{prop:fund_and_absinv}
Let $\tilde{\mathcal{L}}^{\vec{d}}$ be the absorption inverse of $\tilde{\mathcal{L}} $ with respect to $\vec{d}$. Let $\vec{u}$ be a vector in $\Ker \tilde{\mathcal{L}}$ with positive entries $u_i$ such that 
$\sum_i u_i =1$. Additionally, let $D := \diag\{\vec{d}\}$ and $U := (1/\hat{\delta}) \vec{u} \vec{1}^{\T}$, where $\hat{d} = \sum_i (u_i d_i)$. The following relationship holds: 
\begin{equation}\label{eqn:fund_and_absinv}
	(\tilde{\mathcal{L}} + D)^{-1} = U + (I + \tilde{\mathcal{L}}^{\vec{d}}D)^{-1}\tilde{\mathcal{L}}^{\vec{d}} \, .
\end{equation}
\end{proposition}

If $\vec{d} = \vec{\delta}$ and $\tilde{\mathcal{L}} = \mathcal{L} = W-A$, then Propositions \ref{prop:group_inv} and \ref{prop:fund_and_absinv} relate the graph Laplacian matrix $\tilde{\mathcal{L}}(D_\delta,\mathbf{0}) = \mathcal{L}D_\delta^{-1}$ that has the associated absorption-scaled graph $\tilde{G}(D_\delta,\mathbf{0})$ to the continuous-time absorbing Markov chain that is associated with $\mathcal{L}=W - A$ and $\delta_1,\ldots,\delta_n$. We suppose that the spectral radius $\rho(\mathcal{L}^{\vec{\delta}}D_\delta)$ satisfies $\rho(\mathcal{L}^{\vec{\delta}}D_\delta) < 1$ and insert the series expansion $(I + \mathcal{L}^{\vec{\delta}}D_\delta)^{-1} = \sum_{k=0} (-\mathcal{L}^{\vec{\delta}}D_\delta)^k$ into \eqref{eqn:fund_and_absinv} to obtain 
\begin{equation}\label{eqn:series_fund}
	(\mathcal{L} + D_\delta)^{-1} = U + \mathcal{L}^{\vec{\delta}} + \sum_{k=1} (-\mathcal{L}^{\vec{\delta}}D_\delta)^k \mathcal{L}^{\vec{\delta}} \, .   
\end{equation}
For $\epsilon := {\norm{\mathcal{L}^{\vec{\delta}}D_\delta}}\ll 1$, it follows from (\ref{eqn:series_fund}) that 
\begin{equation}\label{eqn:Ld_epsilon}
	 \mathcal{L}^{\vec{\delta}} = (\mathcal{L}+D_\delta)^{-1} - (1/\hat{\delta})u\vec{1}^{\T} + \mathcal{O}(\epsilon) \, .   
\end{equation}
This approximation indicates that when $\epsilon \ll 1$, the entries of $\mathcal{L}^{\vec{\delta}}$ approximate the {differences in the expected times} to absorption between the continuous-time absorbing Markov chain that is associated with $\mathcal{L}$ and $\delta_1,\ldots,\delta_n$ and the Markov chain without absorption that is associated with $(1/\hat{\delta})u\vec{1}^{\T}$. 

To compute the absorption inverse, we use the following proposition.

\begin{proposition}[\cite{jacobsen2018generalized}, Lemma 3 and Theorem 3] \label{prop:compute_abs_inv}
Let $\vec{d}$ be an absorption-rate vector, and let $\vec{u} \in \Ker \tilde{\mathcal{L}} = \Ker {(W-A)}$. Additionally, let $\vec{w} := W_{\dg}$, $\pi := W\vec{u}/(\vec{w}^{\T}\vec{u})$, $D := \diag\{\vec{\delta}\}$, $U := (1/(\vec{d}^{\T}\vec{u}))\vec{u}\vec{1}^{\T}$, and $Z:= W^{-1}Z_0$, where $Z_0 = \left(I-AW^{-1}+\pi \vec{1}^{\T}\right)^{-1}$ is the fundamental matrix that is associated with $AW^{-1}$. We have that
\begin{equation}
	\tilde{\mathcal{L}}^{\vec{d}} = (I-UD)Z(I-DU) \,.
\end{equation}
\end{proposition}

The following proposition relates the absorption inverse $\mathcal{L}^{\vec{\delta}}$ (which is associated with $\tilde{G}(D_\delta,\mathbf{0})$) and {an absorption inverse
of $\tilde{\mathcal{L}}(D_\delta,I)$} (which is associated with $\tilde{G}(D_\delta,I)$). We {prove} this proposition in Appendix \ref{appendix}.

\begin{proposition} \label{prop:L0L1}
Let $\tilde{\mathcal{L}}_1  := \tilde{\mathcal{L}}(D_\delta,I) = (W - A)(W + D_\delta)^{-1}$, and let $\vec{d}_1$ be the diagonal of $D_\delta(W + D_\delta)^{-1}$. Additionally, let $U:= \vec{u}\vec{1}^{\T}/(\vec{\delta}^{\T}\vec{u})$, $U_1 := (W + D_\delta)U$, and $D_1 := D_\delta(W+D_\delta)^{-1}$. We have that
\begin{equation} \label{eqn:L0L1}
	\tilde{\mathcal{L}}_1^{\vec{d}_1} = (W+D_\delta) \mathcal{L}^{\vec{\delta}}  
\end{equation}
and 
\begin{equation}\label{eqn:L1fund}
	(\mathcal{L}+D_\delta)^{-1} = (W+D_\delta)^{-1}\left(U_1 + (I + \tilde{\mathcal{L}}_1^{\vec{d}_1}D_1)^{-1}\tilde{\mathcal{L}_1}^{\vec{d}_1}\right) \,.  
\end{equation}
\end{proposition}

Let $\vec{d}' := \vec{d}_{\mathrm{s}}(D_\delta,I)$ be the absorption-rate vector that is associated with $H = I$. The following proposition relates the absorption inverse $\mathcal{L}^{\vec{\delta}}$ (which is associated with $\tilde{G}(D_\delta,\mathbf{0})$) and the absorption inverse $\mathcal{L}^{\vec{d}'}$ (which is associated with $\tilde{G}(D_\delta,I)$). We {prove} this proposition in Appendix \ref{appendix}.  

\begin{proposition} \label{prop:d0_d1}
Let $\vec{d}' := \vec{d}_{\mathrm{s}}(D_\delta,I) = \vec{w} + \vec{\delta} =  (\omega_1+\delta_1,\ldots, \omega_n + \delta_n)^{\T}$ be the scaled rate vector that is associated with the absorption-scaled graph $\tilde{G}(D_\delta, I)$. With $\mathcal{L} = W - A$, $\alpha := \vec{\delta}^{\T}\vec{u}/(\vec{w}^{\T}\vec{u} + \vec{\delta}^{\T}\vec{u})$, $\pi = W\vec{u}/(\vec{w}^{\T}\vec{u})$, $Z_0 = \left(I - AW^{-1} + \pi \vec{1}^{\T}\right)^{-1}$, and $Z_* = W^{-1}\left(Z_0 - \pi \vec{1}^{\T} \right)$, it follows that
\begin{equation} \label{eqn:0pf}
	\mathcal{L}^{\vec{d}'} = \alpha^2\mathcal{L}^{\vec{\delta}} + \alpha(1-\alpha)\left(\mathcal{L}^{\vec{\delta}}\mathcal{L}Z_*+Z_*\mathcal{L}\mathcal{L}^{\vec{\delta}} \right) +(1-\alpha)^2Z_*   \, . 
\end{equation}
\end{proposition}

Because $Z_0$ is the fundamental matrix of the regular Markov chain with transition-probability matrix $ AW^{-1}$, the entries of $Z_* = W^{-1}(Z_0 - \pi \vec{1}^{\T})$ measure the differences in expected visit times between the Markov chain that is associated with $AW^{-1}$ and the Markov chain that is associated with $\pi\vec{1}^{\T}$. Therefore, if  $\norm{D_\delta}/\norm{\mathcal{L}}\ll 1$, then the matrix $Z_*$ is the analogue of $\mathcal{L}^{\vec{\delta}}$. From \eqref{eqn:0pf}, we see that $\mathcal{L}^{\vec{d}'}$ is a linear combination of terms that include $\mathcal{L}^{\vec{\delta}}$ and $Z_*$, where the corresponding coefficients depend on $\alpha = \vec{\delta}^{\T}\vec{u}/(\vec{w}^{\T}\vec{u} + \vec{\delta}^{\T}\vec{u})$. Additionally, $\alpha \approx 1$ implies that $\mathcal{L}^{\vec{d}'} \approx \mathcal{L}^{\vec{\delta}}$ and $\alpha \approx 0$ implies that $\mathcal{L}^{\vec{d}'} \approx Z_*$.

\section{Examples} \label{sec:examples}

In this section, we apply Algorithms \ref{alg:info_abs_partA} and \ref{alg:info_abs_partB} to three small networks.

\subsection{A three-node network}\label{sec:3nodesEx}

We consider the three-node network (see Figure \ref{ex1}) with adjacency matrix $A$ in (\ref{eqn:A_3nodes}) and node-absorption-rate vector $\vec{\delta} = (\delta_1,\delta_2,\delta_3)^{\T}$. 


\begin{figure}[H]
\begin{center}
        \includegraphics[scale = 0.4]{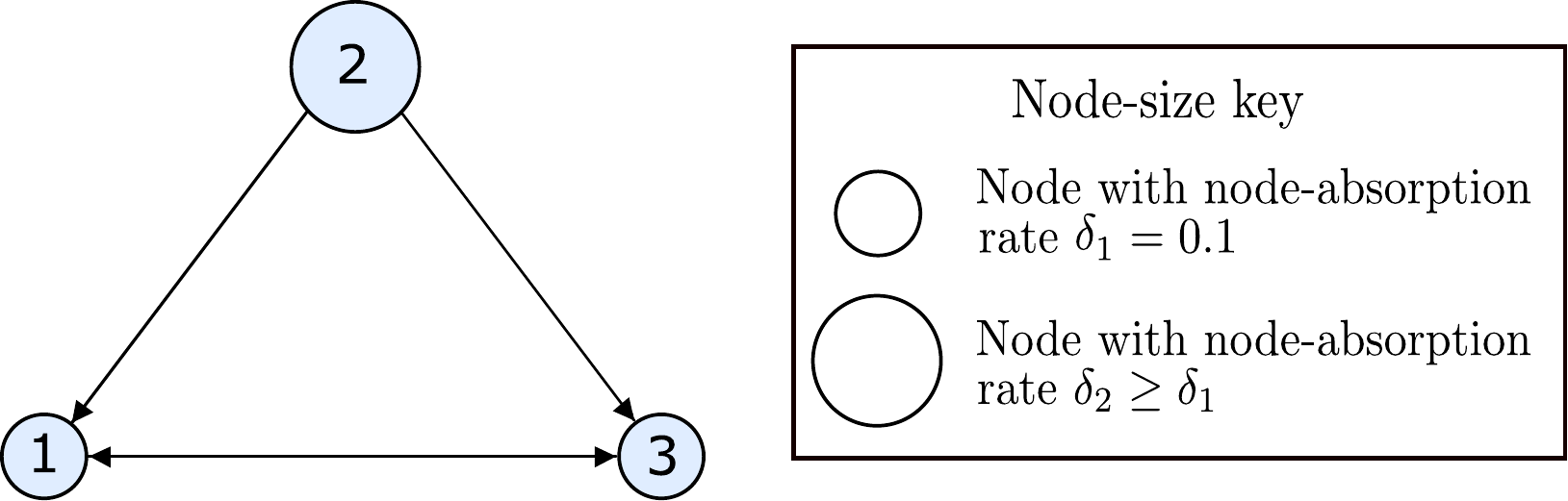}
        \caption{An example three-node network. {The node-absorption rate of node 2 is {greater than or equal to} the node-absorption rates of nodes 1 and 3.}
        }
        \label{ex1}
\end{center}
\end{figure}

Consider the absorption-scaled graph $\tilde{G}(D_\delta,H)$ with $H = \mathbf{0}$ and Algorithm \ref{alg:info_abs_partA} with input $P_l(D_\delta,\mathbf{0},t)$ for a fixed Markov time $t$. In Figure \ref{fig:3nodes}, we show the values of $L(M) = L(M, P_l(D_\delta, \mathbf{0}, 1/20))$ for all five possible partitions $M$ of the three nodes, where we fix $\delta_1 = \delta_3 = 0.1$ and vary $\delta_2$ in the interval $[0.1,1]$. When all of the absorption rates are equal (i.e., when $\delta_2 = 0.1$), we see that $L(M)$ is minimized by the partitions $M = \{\{2\},\{1,3\}\}$ and $M = \{\{1,2,3\}\}$. (See the dashed orange curve.) However, when $\delta_2 > 0.1$, the partition $M = \{\{2\},\{1,3\}\}$ produces a smaller value of the map function (see the solid blue curve); this partition is the output of Algorithm \ref{alg:info_abs_partA}.


\begin{figure}[H]
    \centering
    \includegraphics[scale=0.5]{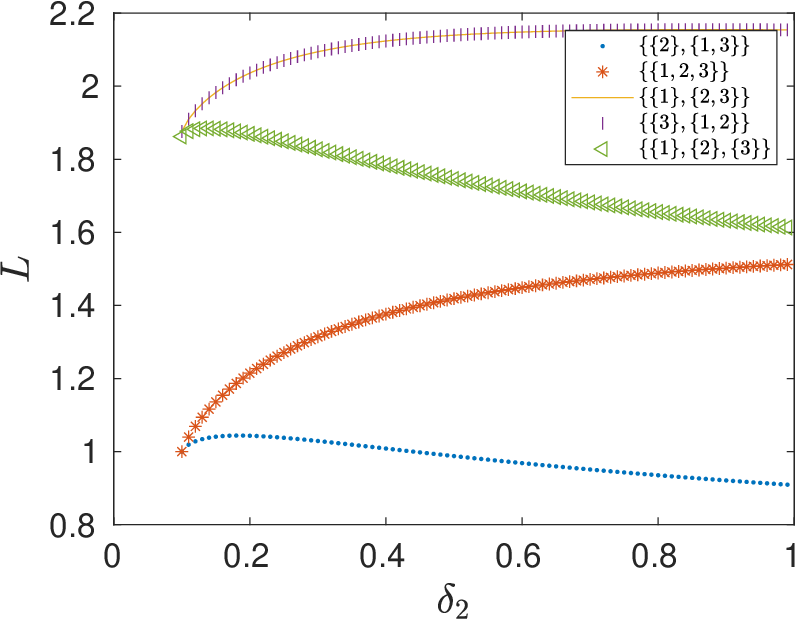}
    \caption{The values of $L(M, P_l(D_\delta, \mathbf{0}, 1/20))$ for all five possible partitions $M$ of 
    the three-node network in Figure \ref{ex1} with $\delta_1 = \delta_3 = 0.1$ and Markov time $t = 1/20$.
    }
    \label{fig:3nodes}
\end{figure}

\subsection{A four-clique network}\label{sec:4cliquesEx}

Consider a network that consists of four planted cliques [see Figure \ref{fig:4comm2struct}(a)], with four nodes each. Additionally, suppose that all of the edge weights are $1$. The 
node-absorption rates of the nodes in the cliques $C_1 := \{1,2,3,4\}$ and $C_3 := \{9,10,11,12\}$ are $\delta_i = 7$ (with $i \in C_1 \cup C_3$), and the node-absorption rates of the nodes in the cliques $C_2 := \{5,6,7,8\}$ and $C_4 := \{13,14,15,16\}$ are $\delta_i = 1$ (with $i \in C_2 \cup C_4$). 

Consider the node-absorption-rate vector $\vec{\delta} = (\delta_1,\ldots, \delta_{16})^{\T}$ and the absorption-scaled graphs $\tilde{G}(D_\delta, \mathbf{0})$ and $\tilde{G}(D_\delta, (3/2) I)$. We use Algorithm \ref{alg:info_abs_partA} with inputs $P_l(D_\delta, \mathbf{0},t)$ and $P_l(D_\delta, (3/2) I,t)$ and Algorithm \ref{alg:info_abs_partB} with inputs $P_e(D_\delta, \mathbf{0},t)$ and $P_e(D_\delta, (3/2) I,t)$ for different Markov times $t$. Arguably, the network structure (i.e., the network topology and the edge weights) on its own favors the partition $M^* := \{C_1,C_2,C_3,C_4\}$ in Figure \ref{fig:4comm2struct}(b). However, the larger absorption rates in the cliques $C_1$ and $C_3$ lead to the partition $M^{**} := \{ \{1\},\{2\},\{3\},\{4\}, C_2, \{9\},\{10\},\{11\},\{12\},C_4\}$ in Figure \ref{fig:4comm2struct}(c).

\begin{figure}[H]
     \centering
     \subfloat[][An example network with four planted cliques]{\includegraphics[width=0.45\textwidth]{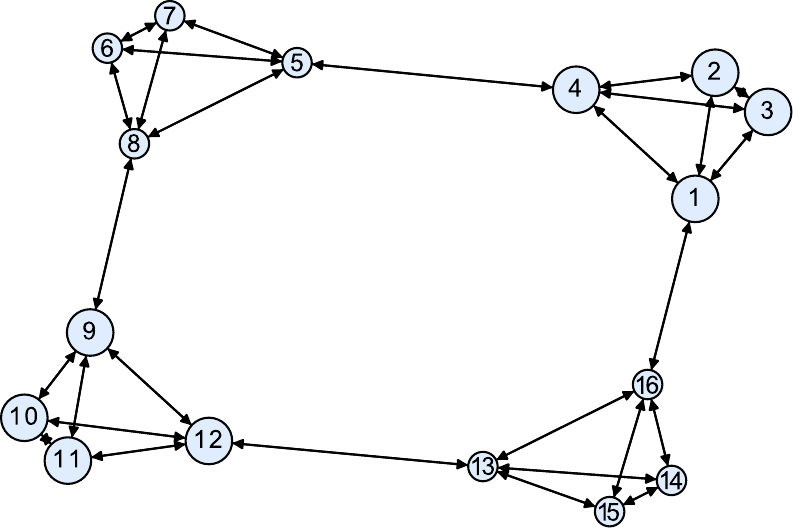}}
    \subfloat[][The partition $M^*$]{\includegraphics[scale=0.55]{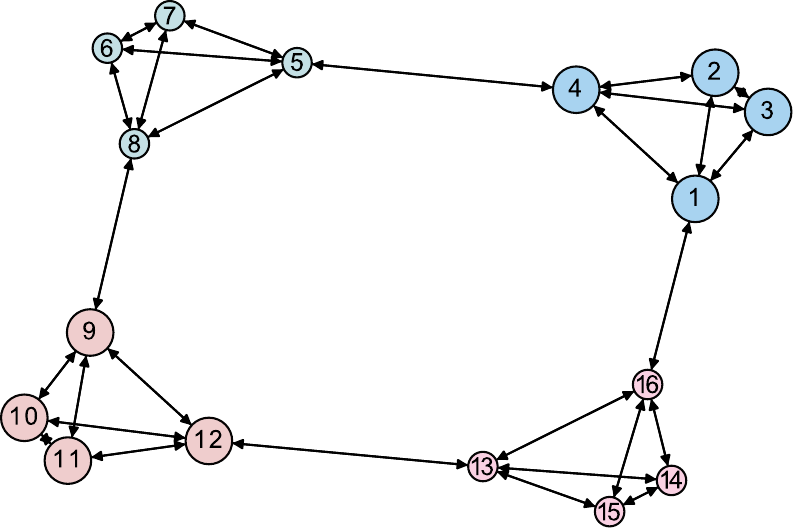}} \\
    \subfloat[][The partition $M^{**}$]{\includegraphics[scale=0.55]{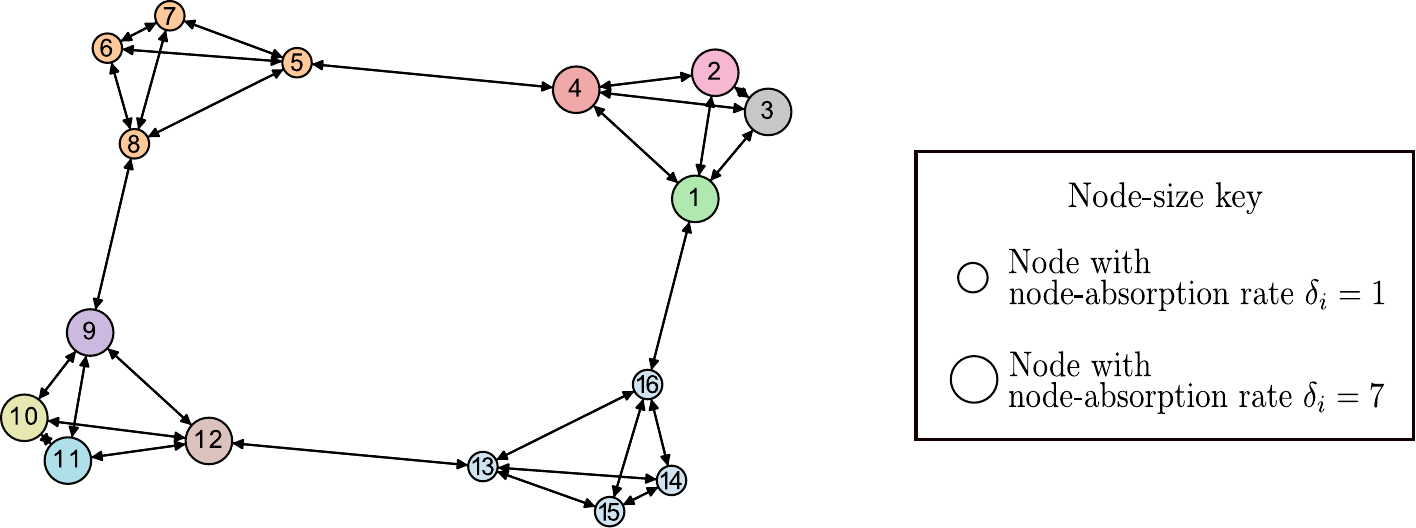}}
     \caption{An example network with four planted cliques and two partitions of its set of nodes. In (a), the node-absorption rates of the large nodes are $\delta_i = 7$ and the node-absorption rates of the small nodes are $\delta_i = 1$. The edge weights are all equal to $1$.
     In (b) and (c), each color indicates a different community. The partition $M^*$ in (b) is the four-clique planted partition, which arises from the network structure (i.e., the network topology and the edge weights). The partition $M^{**}$ in (c) arises from a combination of the network structure and the node-absorption rates.
     }
     \label{fig:4comm2struct}
\end{figure}

Algorithms \ref{alg:info_abs_partA} and \ref{alg:info_abs_partB} with $H = \mathbf{0}$ produce the partition $M^*$ for a smaller range of Markov times than Algorithms \ref{alg:info_abs_partA} and \ref{alg:info_abs_partB} with $H = (3/2)I$. Specifically, Algorithm \ref{alg:info_abs_partA} with the input $P_l(D_\delta,\mathbf{0},t)$ does not produce the partition $M^*$ for any Markov time $t$ that satisfies (\ref{eqn:feasible_alg}) [see Figure \ref{fig:4commMarkovTimes}(a)], whereas Algorithm \ref{alg:info_abs_partA} with the input $P_l(D_\delta,(3/2)I,t)$ produces the partition $M^*$ when $1.47 \lessapprox t \lessapprox 1.75$ [see Figure \ref{fig:4commMarkovTimes}(c)]. Additionally, Algorithm \ref{alg:info_abs_partB} with the input $P_{{e}}(D_\delta,\mathbf{0},t)$ produces $M^*$ when $1.28 \lessapprox t \lessapprox 2.67$ [see Figure \ref{fig:4commMarkovTimes}(b)], whereas Algorithm \ref{alg:info_abs_partB} with the input $P_{{e}}(D_\delta, (3/2)I,t)$ produces $M^*$ when $2.01 \lessapprox t \lessapprox 13.73$ [see Figure \ref{fig:4commMarkovTimes}(d)]. These results are consistent with the fact that the choice $H = (3/2)I$ gives more importance to the edge weights than the choice $H = \mathbf{0}$.

\begin{figure}[H]
     \centering
     \subfloat[][Input $P_l(D_\delta,\mathbf{0},t)$]{\includegraphics[scale=0.4]{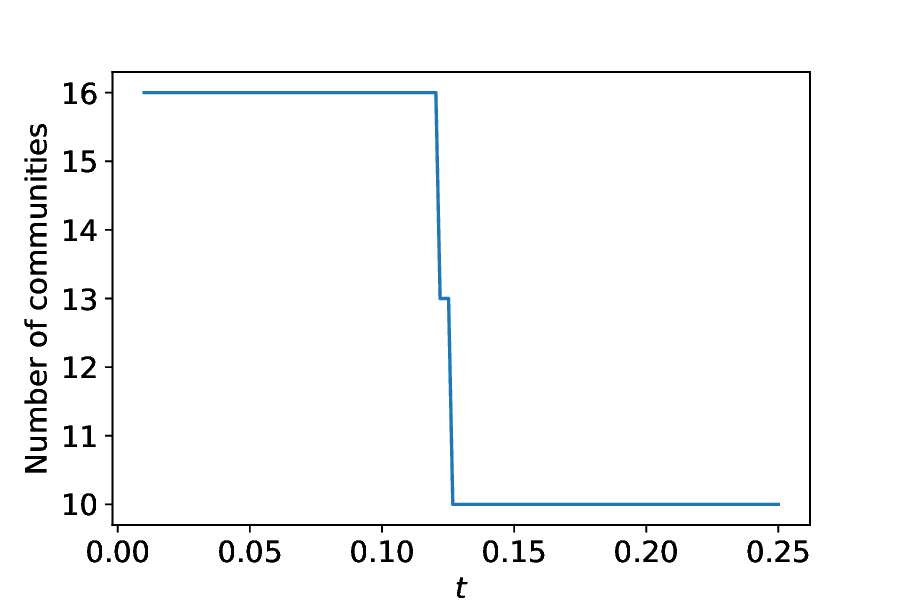}}
     \subfloat[][Input $P_e(D_\delta,\mathbf{0},t)$]{\includegraphics[scale=0.4]{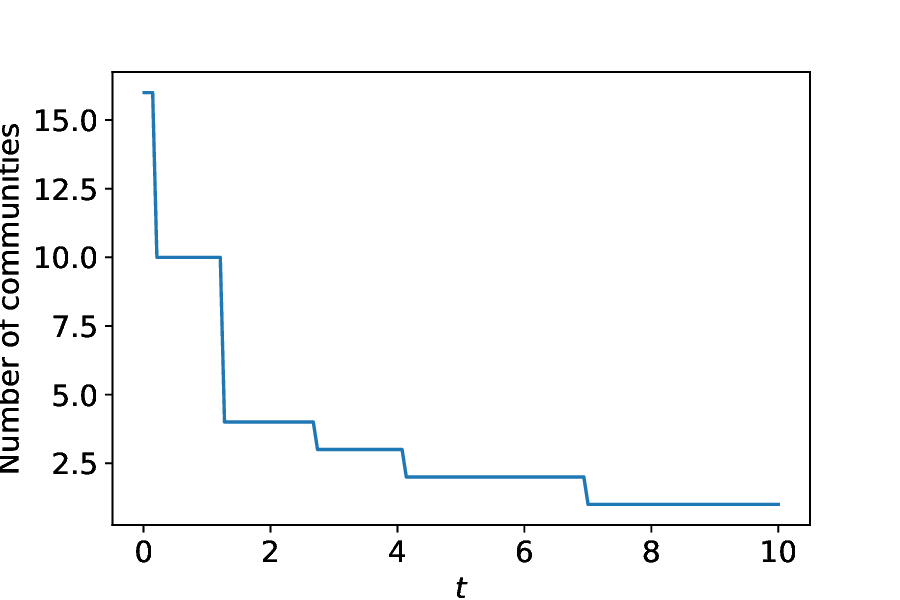}}\\
      \subfloat[][Input $P_l(D_\delta,{(3/2)} I,t)$]{\includegraphics[scale=0.4]{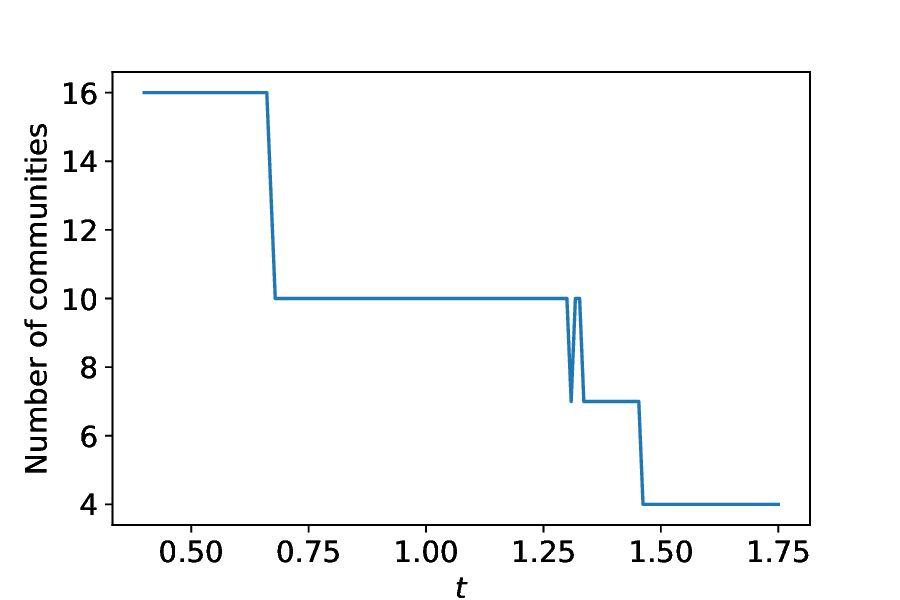}}
    \subfloat[][Input $P_e(D_\delta,{(3/2)} I,t)$]{\includegraphics[scale=0.4]{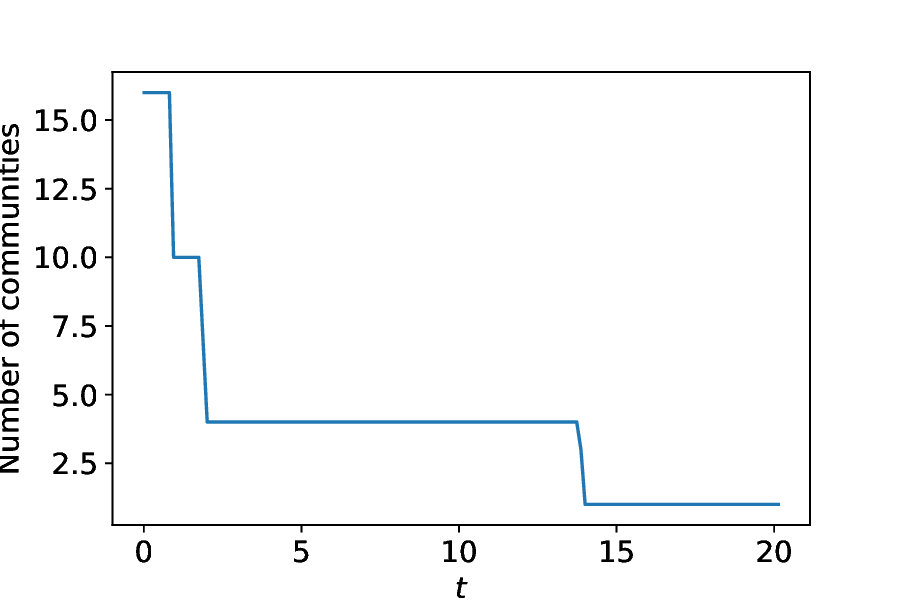}} 
     \caption{The numbers of communities in the partitions that we obtain using Algorithms \ref{alg:info_abs_partA} and \ref{alg:info_abs_partB} with four different inputs. The partitions that consist of four communities {are the same as the partition} $M^*$ in Figure \ref{fig:4comm2struct}(b), and the partitions that consist of ten communities are the same as the partition $M^{**}$ in Figure \ref{fig:4comm2struct}(c). } 
     \label{fig:4commMarkovTimes}
\end{figure}

\subsection{A {square-lattice} network with four different node-absorption rates}\label{sec:gridEx}

We consider a {square-lattice} network with 36 nodes (see Figure \ref{fig:gridEx}). We divide the set of nodes into four {sublattices} (with labels $B_1$, $B_2$, $B_3$, and $B_4$), which we illustrate using nodes of different sizes in Figure \ref{fig:gridEx}. We endow the nodes in {sublattice} $B_1$ with a node-absorption rate of $0.2$, the nodes in {sublattice} $B_2$ with a node-absorption rate of $0.7$, the nodes in {sublattice} $B_3$ with a node-absorption rate of $1.2$, and the nodes in {sublattice} $B_4$ with a node-absorption rate of $1.7$.


\begin{figure}[H]
\begin{center}
        \includegraphics[width=0.5\textwidth]{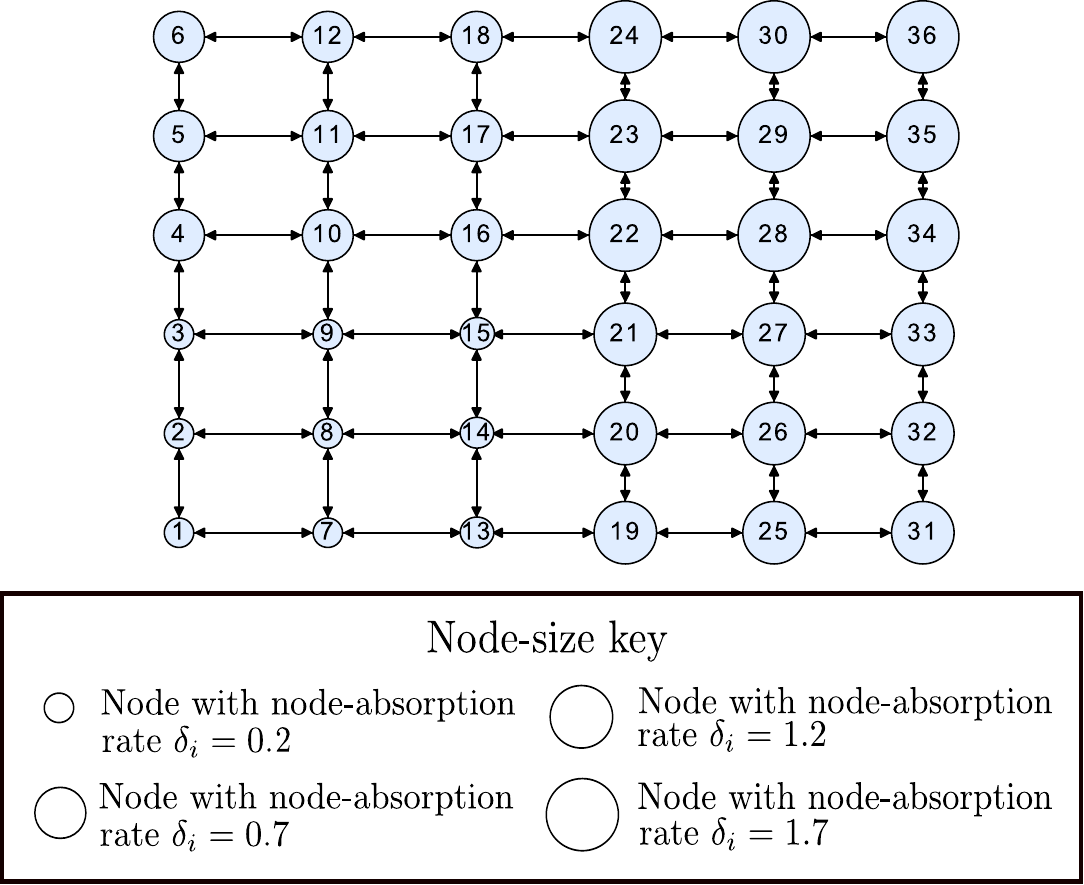}
        \caption{A {square-lattice} network with nodes that have node-absorption rates in the set $\{0.2,0.7,1.2,1.7\}$. 
        }
        \label{fig:gridEx}
\end{center}
\end{figure}

We first look at the community structure that we obtain using Algorithm \ref{alg:info_abs_partA} with the input $P_l(D_\delta,\mathbf{0},t)$ for a Markov time $t$ that satisfies \eqref{eqn:feasible_alg}. In Figure \ref{fig:gridLinearExp}(a), we see that we obtain a partition with 28 communities when $t \gtrapprox 0.0345$. In this partition, $B_1$ is a community and all other communities consist of individual nodes [see Figures \ref{fig:gridLinearExp}(a) and \ref{fig:gridLinearExp}(c)].
By contrast, Algorithm \ref{alg:info_abs_partB} produces {different} communities than Algorithm \ref{alg:info_abs_partA} for {sufficiently large Markov times $t$.} For example, Algorithm \ref{alg:info_abs_partB} with the input $P_e(D_\delta, I, 5.25)$ produces the partition $\{B_1, B_2, B_3, B_4\}$ [see Figure \ref{fig:gridLinearExp}(d)].


\begin{figure}[H]
     \centering
     \subfloat[][Input $P_l(D_\delta, \mathbf{0}, t)$]{\includegraphics[scale=0.5]{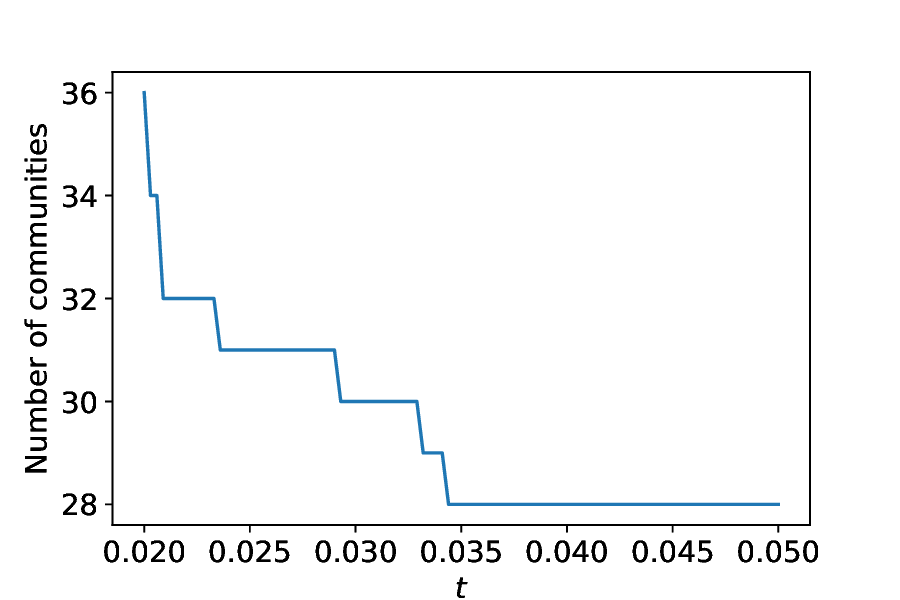}}
     \subfloat[][Input $P_e(D_\delta, I, t)$]{\includegraphics[scale=0.5]{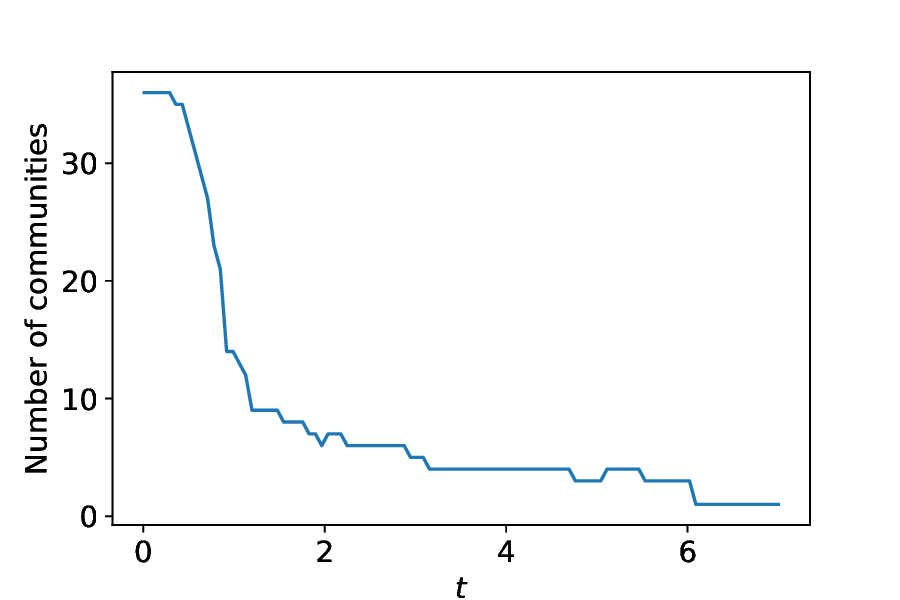}}\\
     \subfloat[][Resulting partition for $P_l(D_\delta, \mathbf{0}, 0.04)$]{\includegraphics[scale=0.38]{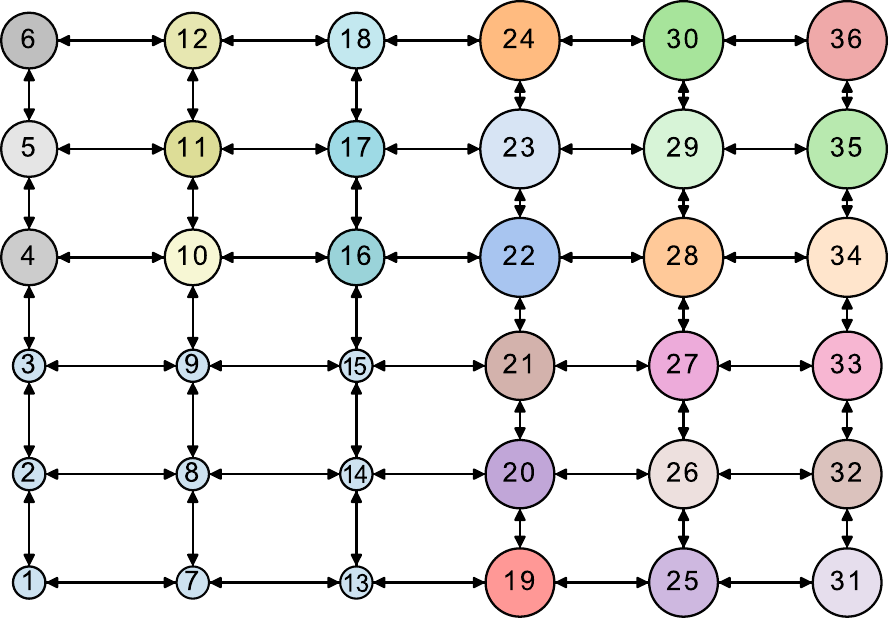}} \quad \quad
      \subfloat[][Resulting partition for $P_e(D_\delta, I, 5.25)$]{\includegraphics[scale=0.39]{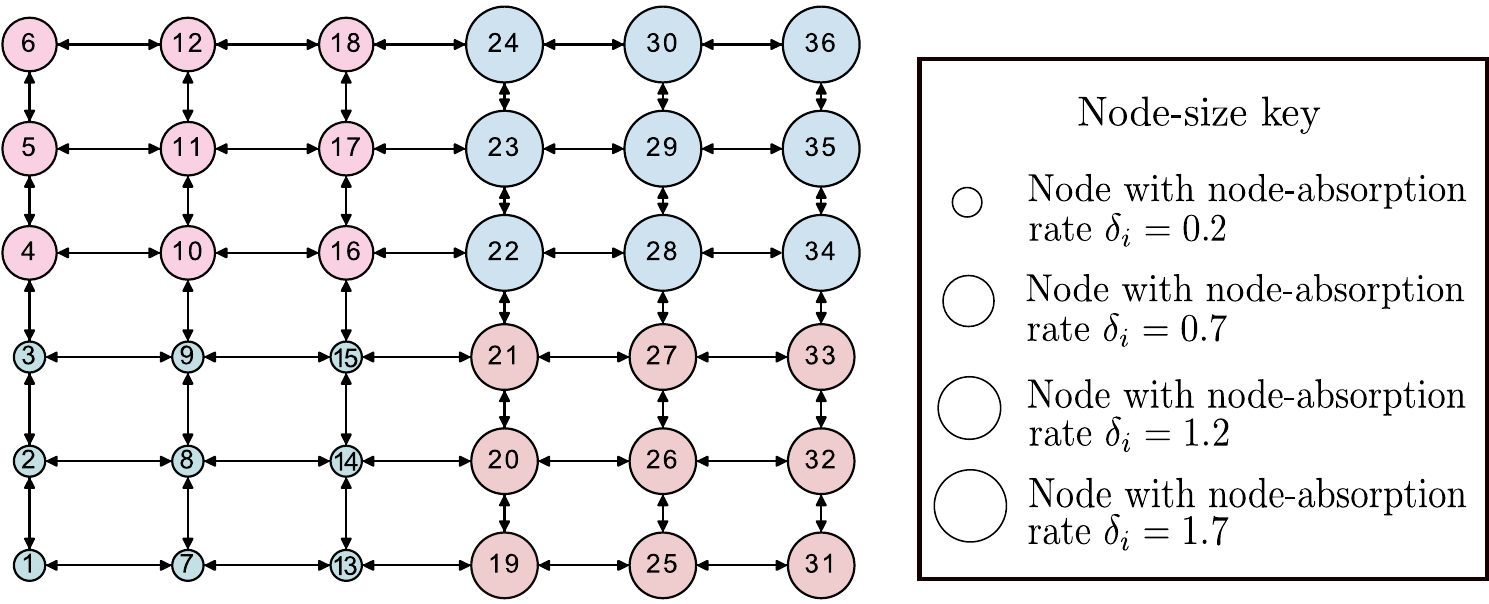}}
     \caption{(a,b) The numbers of communities in the partitions that we obtain using (a) Algorithm \ref{alg:info_abs_partA} with the input $P_l(D_\delta, \mathbf{0}, t)$ and (b) Algorithm \ref{alg:info_abs_partB} with the input $P_e(D_\delta, I, t)$. In (c) and (d), {we show} the partitions that we obtain from Algorithm \ref{alg:info_abs_partA} with the input $P_l(D_\delta, \mathbf{0}, 0.04)$ and Algorithm \ref{alg:info_abs_partB} with the input $P_e(D_\delta, I, 5.25)$, respectively.  
     The colors in panels (c) and (d) indicate community assignments. 
     } 
     \label{fig:gridLinearExp}
\end{figure}

\section{Effective community structure and susceptible--infected--recovered dynamics on networks with ring-lattice communities}\label{sec:small_world}

Inspired by an example in Salath\'e and Jones \cite{salathe2010} about the impact of community structure on disease spread, we explore the impact of the effective community structure of a contact network on the outbreak duration, final size, and outbreak peak of a disease in simulations of a susceptible--infected--recovered (SIR) model of disease spread on the network. {See \cite{kiss2017,pastor2015epidemic} for an introduction to disease dynamics on networks.}

\subsection{An example of Salath\'e and Jones \cite{salathe2010}} \label{sec61}

Salath\'e and Jones \cite{salathe2010} explored the effect of {differences} in community structure on the outbreak duration, final size, and outbreak peak in simulations of an SIR model on networks. In our work, we consider a modified version of one of the examples in \cite{salathe2010}. 
{Salath\'{e} and Jones rewired edges to consider networks with different community structures, and they thereby examined how community structure impacts various aspects of disease outbreaks. Inspired by their exploration, we examine different node-absorption rates (instead of rewiring edges) to consider different community structures, and} 
{we simulate disease dynamics on the resulting networks. {One of the findings of Salath\'{e} and Jones is that outbreak duration depends non-monotonically on {how modular the network is.}}}
We examine whether there is a similar relationship between outbreak duration and the {community structure} that we detect with our adaptation of InfoMap{.} 

{We begin by describing the  example of Salath\'e and Jones{.}  Each stage $s$ {of their example} has an associated network $G^{\mathrm{WS}}_s$ and SIR simulations on that network.}
In stage $s = 1$, Salath\'e and Jones~\cite{salathe2010} considered a network $G^{\mathrm{WS}}_1$ that consists of Watts--Strogatz (WS) small-world graphs \cite{smallworld-scholarpedia}, which yield ``planted communities'' that are connected to each other by ``community bridges'' through edges that are assigned uniformly at random from all possible node pairs. {The set of nodes is {the same} in all stages, but {Salath\'e and Jones rewire the set of edges in each stage.} In stage $s + 1$, Salath\'e and Jones rewired one community bridge of $G^{\mathrm{WS}}_s$ (which is known from stage $s$) into one edge within a planted community in the following manner. First, they 
selected a community bridge $\{i_1,i_2\}$ uniformly at random from all of the community bridges of $G^{\mathrm{WS}}_s$. They then selected a node $i_{k}$ uniformly at random from $\{i_1,i_2\}$ and selected a node $i_3 \neq i_k$ uniformly at random from the planted community of $i_k$. Finally, they obtained the set of edges of $G^{\mathrm{WS}}_{s + 1}$ by removing $\{i_1,i_2\}$ from the union of the edge $\{i_k,i_3\}$ and the set of edges of $G^{\mathrm{WS}}_s$. At each stage $s$, they simulated an SIR process on $G^{\mathrm{WS}}_s$ and recorded the means of the outbreak durations, final sizes, and outbreak peaks. {In all stages, they considered the same time-independent {transmission and recovery intensities} in their SIR simulations.} 

{We also describe the stochastic SIR model that Salath\'e and Jones employed in \cite{salathe2010}. We {simulate this SIR model on networks using a Gillespie algorithm ~\cite{andersson2012stochastic,gillespie1977exact,kiss2017}.} Each node is {either in the susceptible state (i.e., compartment) $S$, the infected state $I$, or the recovered state $R$.} {If a susceptible node has $k$ infected neighbors, then the time that {it} remains susceptible is exponentially distributed with {\emph{transmission intensity}}}\footnote{Our use of the word ``intensity" follows the terminology in~\cite{andersson2012stochastic}.} {$\beta k$. With this distribution, the mean time is $1/(\beta k)$.} The time that an infected node remains infected is exponentially distributed with \emph{recovery intensity} $\gamma$, which yields a mean time of $1/\gamma$.} {Let $|{\rm S}|$ denote the number of susceptible nodes, $|{\rm I}|$ denote the number of infected nodes, and $|{\rm R}|$ denote the number of recovered nodes. Additionally, $|{\rm SI}|$ denotes the number of neighboring node pairs in which one node is susceptible and the others node is infected. {This SIR model is a continuous-time Markov chain {with two possible events:} (1) one susceptible node is infected at a rate $\beta |{\rm SI}|$ {or (2) one infected node} recovers at a rate $\gamma |{\rm I}|$.}} {Two events cannot occur simultaneously.} {In Table \ref{tab:SIR_transitions}, we show the transitions in the numbers of nodes in each state for this model.}

   \begin{table}[H]
   	\centering 
		\begin{tabular}{lr}
		\textbf{Transition} & \textbf{Rate} \\
		\hline \hline
		$|{\rm S}| \rightarrow |{\rm S}| - 1$, $|{\rm I}| \rightarrow |{\rm I}| + 1$, $|{\rm R}| \rightarrow |{\rm R}|$  & \qquad $\beta |{\rm SI}|$ \\
		$|{\rm S}| \rightarrow |{\rm S}|$, $|{\rm I}| \rightarrow |{\rm I}| - 1$, $|{\rm R}| \rightarrow |{\rm R}| + 1$ & \qquad ${\gamma} |{\rm I}|$ \\
		\hline \hline
		\end{tabular}
	\caption{Transitions {in the numbers of nodes in each compartment in the examined SIR model.}}
	\label{tab:SIR_transitions}
   \end{table}


In each stage $s$, Salath\'e and Jones computed the value of modularity of the partition of the network $G^{\mathrm{WS}}_s$. They observed that these modularity values increase with $s$. This reflects {a sparsification of the edges between the planted communities due to the edge-rewiring process.}


\subsection{Increasing node-absorption rates of bridging nodes instead of removing community bridges} \label{following}

\subsubsection{{Ring-lattice graphs as planted communities}}\label{sec:network_SIR}

We study an example that plays a similar role to the example of Salath\'e and Jones~\cite{salathe2010}. In our example, setting the absorption rates of bridge nodes to larger values than the absorption rates of other nodes is analogous to removing community bridges. Unlike in the example of Salath\'e and Jones, our example uses the same network at each stage and we change the absorption rates of specific bridging nodes instead of rewiring community bridges. Because the network is the same in all stages, maximizing modularity or using the standard InfoMap algorithm cannot reveal how effective community structure changes as we change the node-absorption rates, which correspond to the recovery intensities of a disease. However, our adaptations of InfoMap are designed for such situations. The values that we obtain for the associated map function play the role of the modularity values in \cite{salathe2010}{. They reflect an effective sparsification between planted communities due to increasing the} node-absorption rates of some bridging nodes.  
{This effective sparsification between communities, which entails a corresponding increased isolation of communities from each other, is analogous to the more literal sparsification in \cite{salathe2010}.}

{We use Algorithm \ref{alg:generating_G} to generate the network $G^{\mathrm{RL}}$ (which is inspired by the example in \cite{salathe2010} that we described in Section \ref{sec61})
on which we simulate an SIR process. [See Figure \ref{fig:scaled-down}(a) for an example of a subgraph of a network that we generate using Algorithm \ref{alg:generating_G}.] We generate a graph $G^{\mathrm{RL}}$ using Algorithm \ref{alg:generating_G} with $N_{\mathrm{WS}} = 20$ directed and unweighted ring-lattice subgraphs $G_1,\ldots, G_{20}$ (the planted communities) of size $n_{\mathrm{WS}} = 12$ (i.e., with 12 nodes each) that are connected by community bridges (which we place between node pairs that we choose uniformly at random from all possible pairs with nodes in distinct planted communities). Each node of $G^{\mathrm{RL}}$ has $k_{\mathrm{WS}} = 6$ neighbors in its planted community. We choose a small size for the ring-lattice graphs to visualize the effective community structure of one of the 20 ring-lattice subgraphs [see Figures \ref{fig:Stages}(c,d)]. Each of these ring-lattice subgraphs is a one-dimensional lattice with periodic boundary conditions and additional local connections, which we specify in Algorithm \ref{alg:generating_G}. In the context of the analogy with the example in \cite{salathe2010}, each ring-lattice subgraph in our example is a directed WS network in which all edges are bidirectional (i.e., each edge is reciprocated) and the edge-rewiring probability is $0$ \cite{newman2000models}. We use an edge-rewiring probability of $0$ because we seek to examine the effects of changing absorption rates instead of the effects of rewiring edges (which was explored in \cite{salathe2010}). {The degree of a node {of} a network $G^{\textrm{RL}}$ {that we generate} with Algorithm \ref{alg:generating_G} is the random variable $k_{\textrm{WS}} + N_{\textrm{bridges}}${, where $N_{\textrm{bridges}}$ is the number of community bridges that are adjacent to the node. This random variable follows a binomial distribution with parameters $n_{\textrm{WS}}(N_{\textrm{WS}} - 1)$ and $2/(n_{\textrm{WS}}(N_{\textrm{WS}} - 1))$. The mean degree of $G^{\textrm{RL}}$ is $k_{\textrm{WS}} + 2$, and the variance of the degree is $2\left[1 - 2/(n_{\textrm{WS}}(N_{\textrm{WS}} - 1))\right]$. For the graph $G^{\textrm{RL}}$, the standard InfoMap algorithm yields the partition that consists of the planted communities $G_1, \ldots, G_{N_{\textrm{WS}}}$.}}

We use a similar process (see Algorithm \ref{alg:generating_G}) as in \cite{salathe2010} to generate the network $G^{\mathrm{RL}}$ on which we simulate an SIR process. We generate a graph $G^{\mathrm{RL}}$ using Algorithm \ref{alg:generating_G} with $N_{\mathrm{WS}} = 20$ directed and unweighted ring-lattice subgraphs $G_1,\ldots, G_{20}$ (the planted communities) of size $n_{\mathrm{WS}} = 12$ (i.e., with 12 nodes each) that are connected by community bridges (which we place between node pairs that we choose uniformly at random from all possible pairs with nodes in distinct planted communities). Each node of $G^{\mathrm{RL}}$ has $k_{\mathrm{WS}} = 6$ neighbors in its planted community. We choose a small size for the ring-lattice graphs to visualize the effective community structure of one of the 20 ring-lattice subgraphs [see Figures \ref{fig:Stages}(c,d)]. Each of these ring-lattice subgraphs is a one-dimensional lattice with periodic boundary conditions and additional local connections, which we specify in Algorithm \ref{alg:generating_G}. In the context of the analogy with the example in \cite{salathe2010}, each ring-lattice subgraph in our example is a directed WS network in which all edges are bidirectional (i.e., each edge is reciprocated) and the edge-rewiring probability is $0$ \cite{newman2000models}. We use an edge-rewiring probability of $0$ because we seek to examine the effects of changing absorption rates instead of the effects of rewiring edges (which was explored in \cite{salathe2010}).

\begin{algorithm}[H]
\caption{Generation of the network $G^{\mathrm{RL}}$ on which we simulate an SIR process.
}
\label{alg:generating_G}
\begin{algorithmic}[1]
 \renewcommand{\algorithmicrequire}{Input:}
 \renewcommand{\algorithmicensure}{Output:}
 \REQUIRE Positive integers $n_{\mathrm{WS}}$ and $N_{\mathrm{WS}}$; an even positive integer $k_{\mathrm{WS}}$.
   
 \medskip
   
    \ENSURE {An $N$-node} directed and unweighted graph $G^{\mathrm{RL}}$, which we partition into $N_{\mathrm{WS}}$ ring-lattice subgraphs $G_1, \ldots, G_{N_{\mathrm{WS}}}$, which each have size $n_{\mathrm{WS}}$. The graph $G^{\mathrm{RL}}$ has $n_{\mathrm{WS}} \times N_{\mathrm{WS}}$ edges between nodes from distinct ring-lattice subgraphs. Each node of $G^{\mathrm{RL}}$ has $k_{\mathrm{WS}}$ neighbors in its corresponding ring-lattice subgraph. 
    
    \medskip \medskip
	\STATE Define a ring-lattice subgraph $G_1$ with the set $\{1,\ldots,n_{\mathrm{WS}}\}$ of nodes, where the neighbors of each node $i$ are the nodes $i \pm {\ell}$ (mod $n_{\mathrm{WS}}$), with ${\ell} \in \{1,\ldots, k_{\mathrm{WS}}/2\}$. The edges of this subgraph are bidirectional. (In other words, if $(i_1, i_2)$ is an edge of $G_1$, then so is $(i_2,i_1)$.) 
    \medskip    
    \STATE For ${\varphi} \in \{2,\ldots,N_{\mathrm{WS}}\}$, define a ring-lattice subgraph $G_{{\varphi}}$ with the set $\{({\varphi} - 1)\cdot n_{\mathrm{WS}} +1, \ldots, {\varphi}\cdot n_{\mathrm{WS}}\}$ of nodes, where $G_{{\varphi}}$ is isomorphic to $G_1$. (The subgraphs $G_1, \ldots, G_{N_{\mathrm{WS}}}$ are ``planted communities''.)
    \medskip
    \STATE Select $n_{\mathrm{WS}}\times N_{\mathrm{WS}}$ pairs $\{i_1,i_2\}$ of nodes uniformly at random from all node pairs that belong to distinct ring-lattice graphs. Add bidirectional edges between the nodes of each pair. (The pair $\{i_1,i_2\}$ is a {``community bridge''}, and the nodes $i_1$ and $i_2$ {are} ``bridging nodes''.)  
    \medskip
    \STATE Set all of the edge weights to $1$. (That is, we use an unweighted network.)
    \medskip
	
\end{algorithmic}
\end{algorithm}

\subsubsection{{Duration, final size, and peak of SIR outbreaks}}\label{sec:disease_quant}

We use Algorithm \ref{alg:functional} to run SIR simulations on the graph $G^{\mathrm{RL}}$ for different node-absorption {rates} {(which correspond to the recovery intensities)} and transmission {intensities}. {We refer to a set of node-absorption rates $\delta_1, \ldots, \delta_N$ and transmission {intensities} $\beta_1, \ldots, \beta_N$ of the {$N$}} nodes of $G^{\textnormal{RL}}$ as a \emph{parameter configuration}{.} {In steps 1 and 2 of Algorithm \ref{alg:functional}, we recursively {determine} parameter configurations. We refer to each step of this {recursive process} as a \emph{stage}.}  
{For the parameter configurations in two consecutive stages of Algorithm \ref{alg:functional},} increasing the node-absorption rates of two bridging nodes from $\delta_{*}$ to $\delta_{**}$ is analogous to removing a community bridge in \cite{salathe2010} and increasing the transmission {intensities} of other nodes (which we call ``balancing nodes'') from $\beta_{*}$ to $\beta_{**}$ is analogous to adding edges within a community in \cite{salathe2010}. {In Figure \ref{fig:scaled-down}, we show example parameter configurations in two stages of Algorithm \ref{alg:functional} for a 
{{subgraph} of a network {that we generate} with Algorithm \ref{alg:generating_G}.}

\begin{algorithm}[H]
\caption{SIR simulations on a network $G^{\mathrm{RL}}$ (which we generate {with} Algorithm \ref{alg:generating_G}) for different node-absorption {rates} {(i.e., recovery intensities)} and transmission {intensities}. 
}
\label{alg:functional}
\begin{algorithmic}[1]
 \renewcommand{\algorithmicrequire}{Input:}
 \renewcommand{\algorithmicensure}{Output:}
 \REQUIRE A graph $G^{\mathrm{RL}}$ that is the output of Algorithm \ref{alg:generating_G} with inputs $n_{\mathrm{WS}}$ {(the size of each planted community)}, $N_{\mathrm{WS}}$ {(the number of planted communities)}, and $k_{\mathrm{WS}}$ {(the number of neighbors of a node within its planted community).} 
 {(Given these parameters, the graph $G^{\mathrm{RL}}$ has $n_{\mathrm{WS}} \times N_{\mathrm{WS}}$ bidirectional edges (i.e., {community bridges}) that connect nodes from distinct planted communities.)}
 {Node-absorption rates} $\delta_*$ and $\delta_{**}$, with $\delta_{**} > \delta_*$. {(They correspond to the {recovery intensities}.)} Transmission {intensities} $\beta_*$ and $\beta_{**}$, with $\beta_{**} > \beta_*$. Positive integers $N_{\rm S}$ and $N_{\mathrm{sim}}$. 
 \medskip
    \ENSURE The means of the outbreak durations, final sizes, and outbreak peaks of $N_{\mathrm{sim}}$ SIR simulations for each of the $N_{\rm S}$ parameter configurations that are defined by steps 1 and 2 of this algorithm. (A parameter configuration consists of the node-absorption {rates} and transmission {intensities} of the nodes of $G^{\mathrm{RL}}$.)
    \medskip \medskip
	\STATE \underline{{{Determine} the parameter configuration in stage 1:}} Set the node-absorption rate of each node of $G^{\mathrm{RL}}$ to $\delta_*$, and set the transmission {intensity} of each node of $G^{\mathrm{RL}}$ to $\beta_*$. 
    \medskip
    \STATE  \underline{{{Determine} the parameter configuration in stage $s + 1$} for $s \in \{ 1, \ldots, N_{\rm S} - 1\}$:} Once we have determined the {parameter configuration in} stage $s$, select a {bridging-node} pair $\{i_1, i_2\}$ of $G^{\mathrm{RL}}$ uniformly at random from the set of pairs of bridging nodes that have node-absorption rate $\delta_*$. Set the node-absorption rates of nodes $i_1$ and $i_2$ to $\delta_{**}$.
    Additionally, select a node ${j}_k$ uniformly at random from the ring-lattice subgraph that is associated with $i_k$ for $k \in \{1,2\}$. (We require that ${j}_k$ is distinct from $i_k$ and that it is not a neighbor of $i_k$ for $k \in \{1,2\}$.) Set the transmission {intensities} of ${j}_1$ and ${j}_2$ to $\beta_{**}$. 
    \medskip 	  
    \STATE \underline{Run SIR simulations}: {Run $N_{\mathrm{sim}}$ simulations of the SIR model on the graph $G^{\mathrm{RL}}$ with transmission {intensities} $\beta \in \{\beta_*,\beta_{**}\}$ and {recovery intensities} $\gamma \in \{\gamma_{*},\gamma_{**}\}$ that correspond to the parameter configuration in each stage $s \in \{ 1, \ldots, N_{\rm S}\}$.} {The {recovery intensities} $\gamma_*$ and $\gamma_{**}$ equal the associated node-absorption rates $\delta_{*}$ and $\delta_{**}$.} To run these simulations, we use a Gillespie algorithm~\cite{gillespie1977exact}. Record the means of the outbreak durations, the final sizes, and the outbreak peaks of the simulations.
	 \medskip
\end{algorithmic}
\end{algorithm}

\begin{figure}[H]
    \centering
    \subfloat[][Parameter configuration in stage 1]{\includegraphics[scale=0.38]{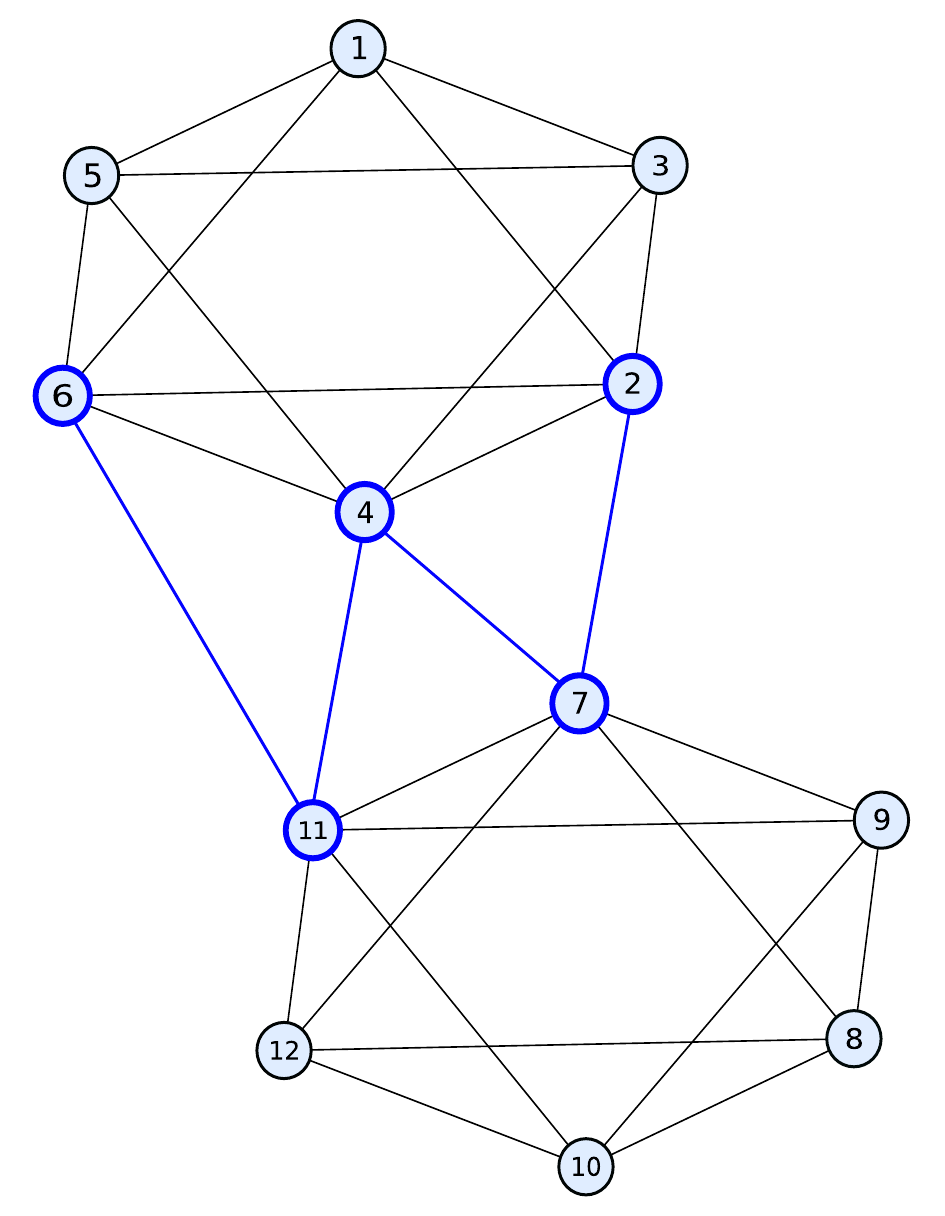}}
    \subfloat[][Parameter configuration in stage 2]{\includegraphics[scale=0.38]{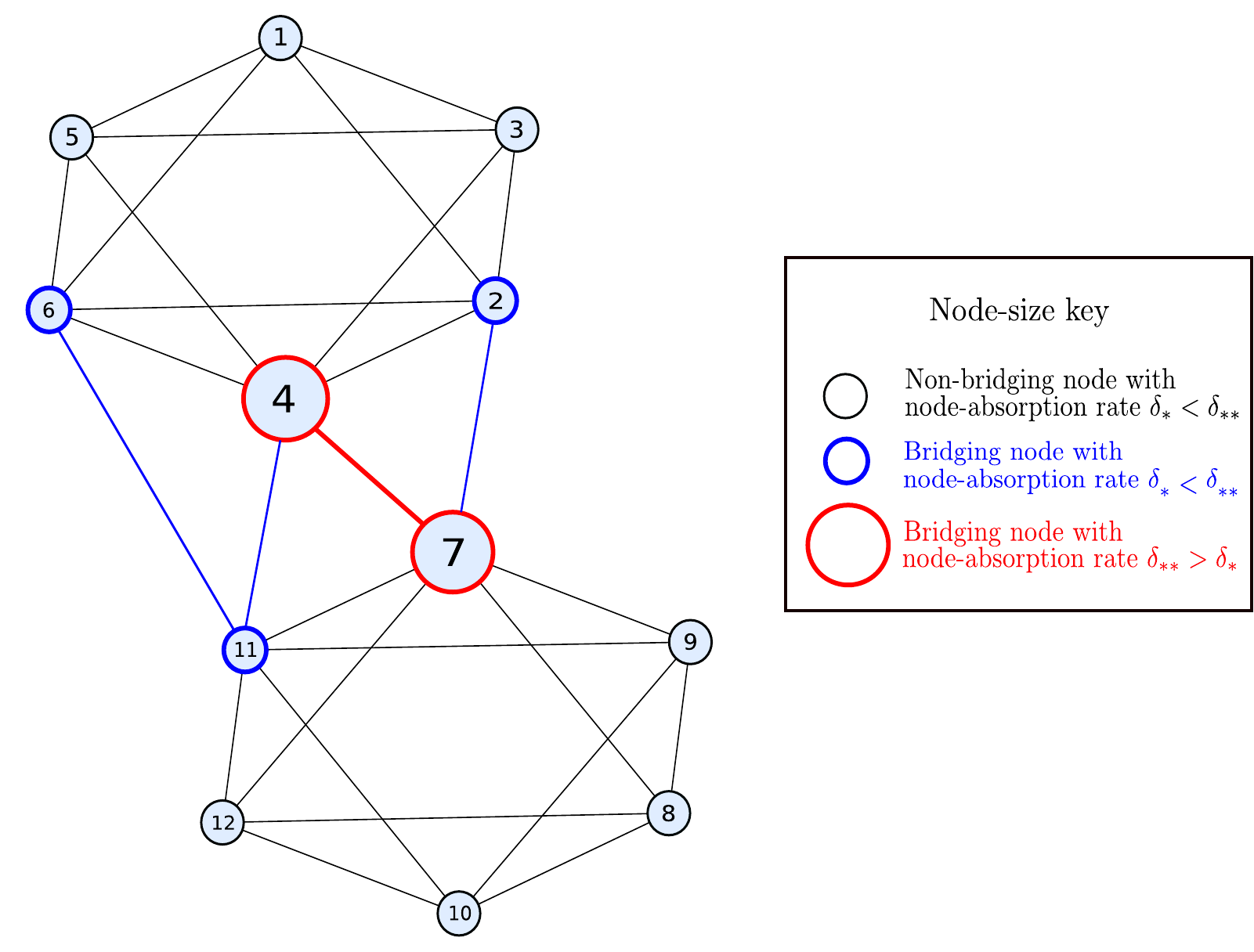}}
    \caption{{Illustration of a {subgraph of a network} that we generate with Algorithm \ref{alg:generating_G} and two stages of Algorithm \ref{alg:functional} for that network. We show the parameter configurations in stages 1 and 2 of Algorithm \ref{alg:functional} for a network $G^{\textrm{RL}}$ from Algorithm \ref{alg:generating_G} with $n_{\textrm{WS}} = 6$, $N_{\textrm{WS}} = 2$, and $k_{\textrm{WS}} = 4$. In this figure, we show only four community bridges (the blue segments) of the {$n_{\textrm{WS}}\times N_{\textrm{WS}} = 12$ community bridges {of $G^{\rm RL}$} between the planted communities $G_1 = \{1,\ldots,6\}$ and $G_2 = \{7,\ldots,12\}$.} (a) In the parameter configuration in stage 1, all 12 nodes have the same node-absorption rate $\delta_{*}$ and the same transmission intensity $\beta_{*}$. 
  (b) In the parameter configuration in stage 2, the bridging nodes 4 and 7 have the same node-absorption rate $\delta_{**}  >\delta_{*}$ and all other nodes have the node-absorption rate $\delta_{*}$. In this stage, nodes 1 and 10 (which are not in the neighborhoods of nodes 4 and 7) have the same transmission intensity $\beta_{**} > \beta_{*}$ and all other nodes have the transmission intensity $\beta_*$.} }
    \label{fig:scaled-down}
\end{figure}

{For each node of $G^{\textnormal{RL}}$, the recovery intensity $\gamma$ equals the associated node-absorption rate {$\delta$}. {The} recovery {intensities} $\gamma_{*}$ and $\gamma_{**}$ {are} $\delta_*$ and $\delta_{**}$, respectively.} We choose the transmission intensity $\beta_{**}$ to compensate for the decrease in new infections that occur {due to} the {increase of the absorption rates of the chosen bridging nodes.} We estimate the decrease in the basic reproduction number $\R_0$ by calculating $\langle k \rangle \beta_*/{\gamma}_* - \langle k \rangle \beta_*/{\gamma}_{**} $ (using an approximation that is similar to one in \cite{newman2002spread}), where $\langle k \rangle$ is the mean degree of the bridging nodes. We compensate for the decrease in infections by choosing $\beta_{**}$ such that $\langle k \rangle \beta_{**}/{\gamma}_* = \langle k \rangle \beta_*/{\gamma}_* + \alpha \left(\langle k \rangle \beta_*/{\gamma}_* - \langle k \rangle \beta_*/{\gamma}_{**} \right)$, which estimates the new infections that arise from balancing nodes, where $\alpha$ is a tuning parameter that we use to preserve the value of the basic reproduction number. This yields

\begin{equation}\label{eqn:compensation}
	\beta_{**} = \beta_* + \alpha {\gamma}_* \beta_*\left(\frac{1}{{\gamma}_*} - \frac{1}{{\gamma}_{**}}\right) \, .
\end{equation}

{We produce {parameter configurations in $N_{\rm S} = 68$ stages} {[}with $\delta_* = 0.2$, $\delta_{**} = 1$, $\beta_{*} = 0.125$, $\alpha = 0.1$, and the corresponding value of $\beta_{**}$ that we obtain using \eqref{eqn:compensation}{]}.} {For each parameter configuration, we run} $N_{\textnormal{sim}} = 1000$ simulations of SIR dynamics on $G^{\mathrm{RL}}$ (see Algorithm \ref{alg:functional}). In Figures \ref{fig:FunctionalQuant}(a)--(c), we show the mean outbreak duration, final size, and outbreak peak for {the parameter configuration {in} each of the 68 stages.} The qualitative behavior of these epidemiological quantities is consistent with the observations in the example of Salath\'e and Jones \cite{salathe2010}. Specifically, the mean final outbreak size and the mean outbreak peak decrease monotonically as we increase the stage number [see Figures \ref{fig:FunctionalQuant}(b,c)]. Additionally, the mean outbreak duration peaks at intermediate stages [see Figure \ref{fig:FunctionalQuant}(a)]. 

In Figure \ref{fig:FunctionalQuant}(d), we show the value of the map function $L(M_0) = L\left(M_0, P_e(D_\delta^{(s)}, \mathbf{0}, t = 0.025)\right)$ for the {partition $M_0$ that consists of the planted communities of $G^{\mathrm{RL}}$}, {where the node-absorption rates in the diagonal {entries} of $D_\delta^{(s)}$ {correspond to} the parameter configuration in stage $s$ of Algorithm \ref{alg:functional}.} (We justify the choice of Markov time $t = 0.025$ in Section \ref{next}.) {This map function, which is associated with our adaptation of InfoMap in Algorithm \ref{alg:info_abs_partB}, captures the effect on community structure of changes in node-absorption rates. The standard map function cannot capture these effects.} In Figure \ref{fig:FunctionalQuant}(d), we observe that the value of the map function $L(M_0)$ decreases with the stage number. Consequently, we conclude from Figure \ref{fig:FunctionalQuant} that the map function $L(M_0)$ decreases as one increases the number of bridging nodes with {the} large node-absorption rate 
$\delta_{**}$. ({For the parameter configuration in stage $s$}, there are $2s$ bridging nodes with {the} large node-absorption rate.) This reflects {an increased isolation of the planted communities.} The monotonicity of $L(M_0)$ is analogous to the increase of modularity that Salath\'e and Jones \cite{salathe2010} obtained for {their} planted partition as they removed community bridges.

\begin{figure}[H]
    \centering
    \subfloat[]{\includegraphics[scale = 0.3]{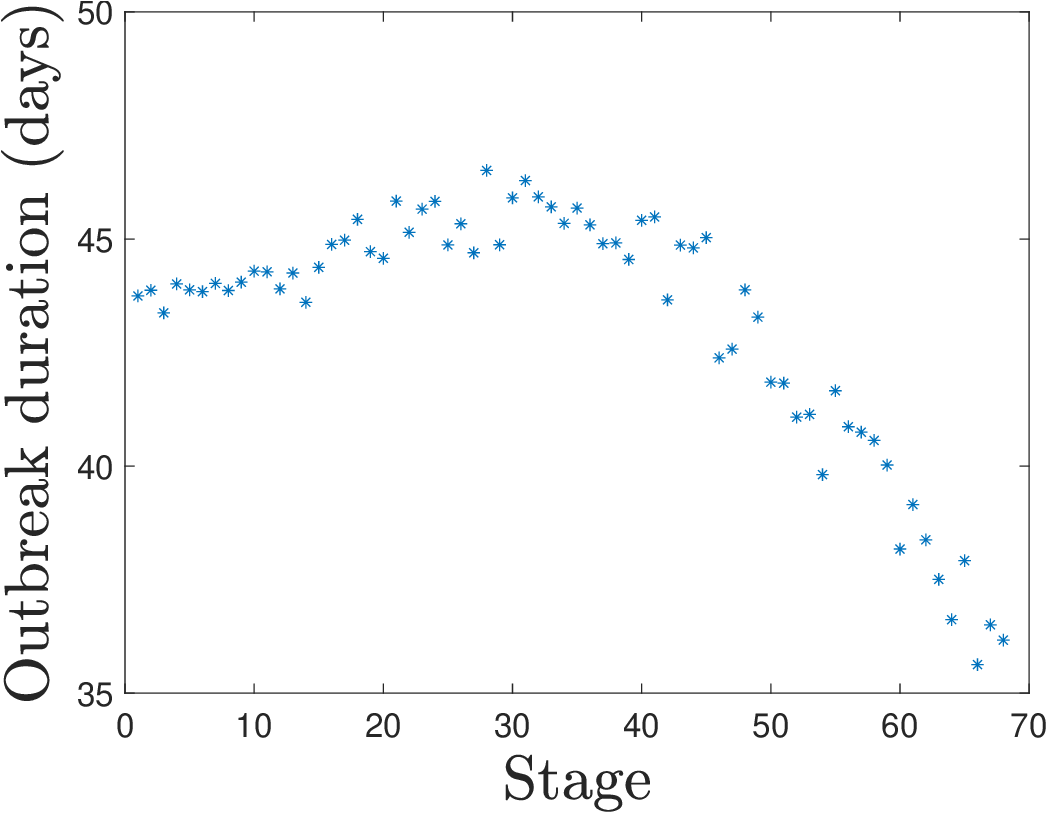}} \quad 
    \subfloat[]{\includegraphics[scale = 0.3]{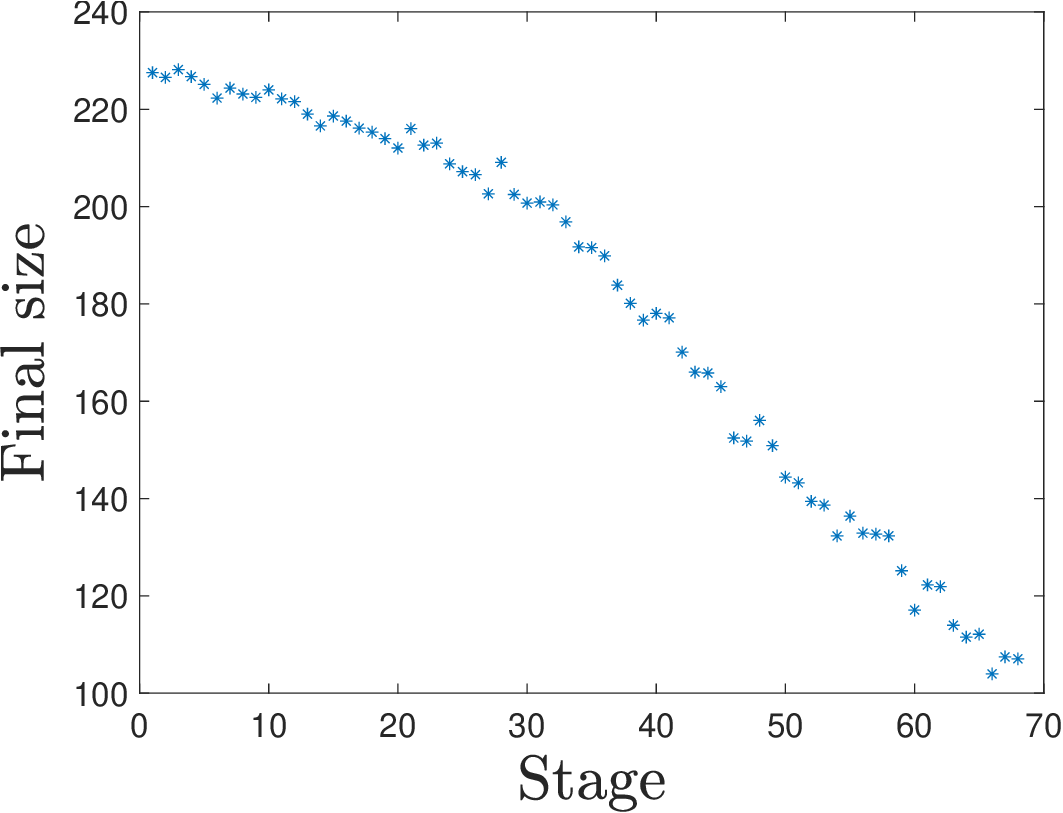}}\\
    \subfloat[]{\includegraphics[scale =0.3]{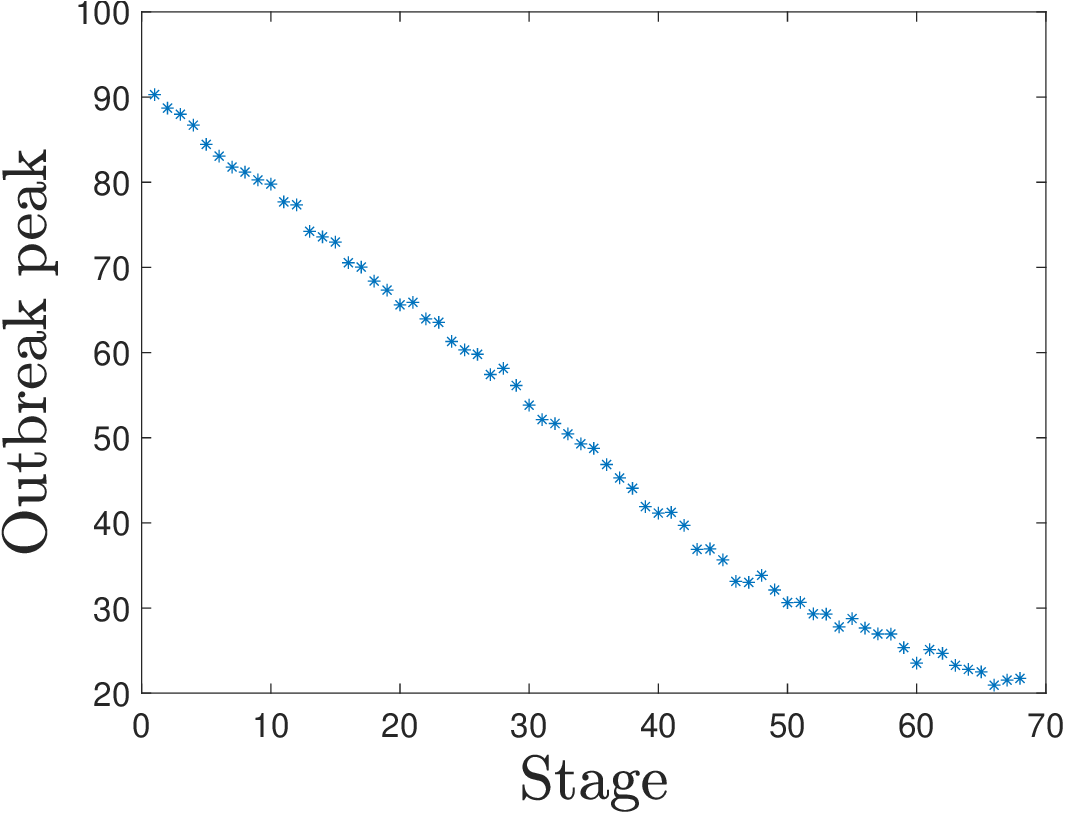}}  \quad
    \subfloat[]{\includegraphics[scale = 0.3]{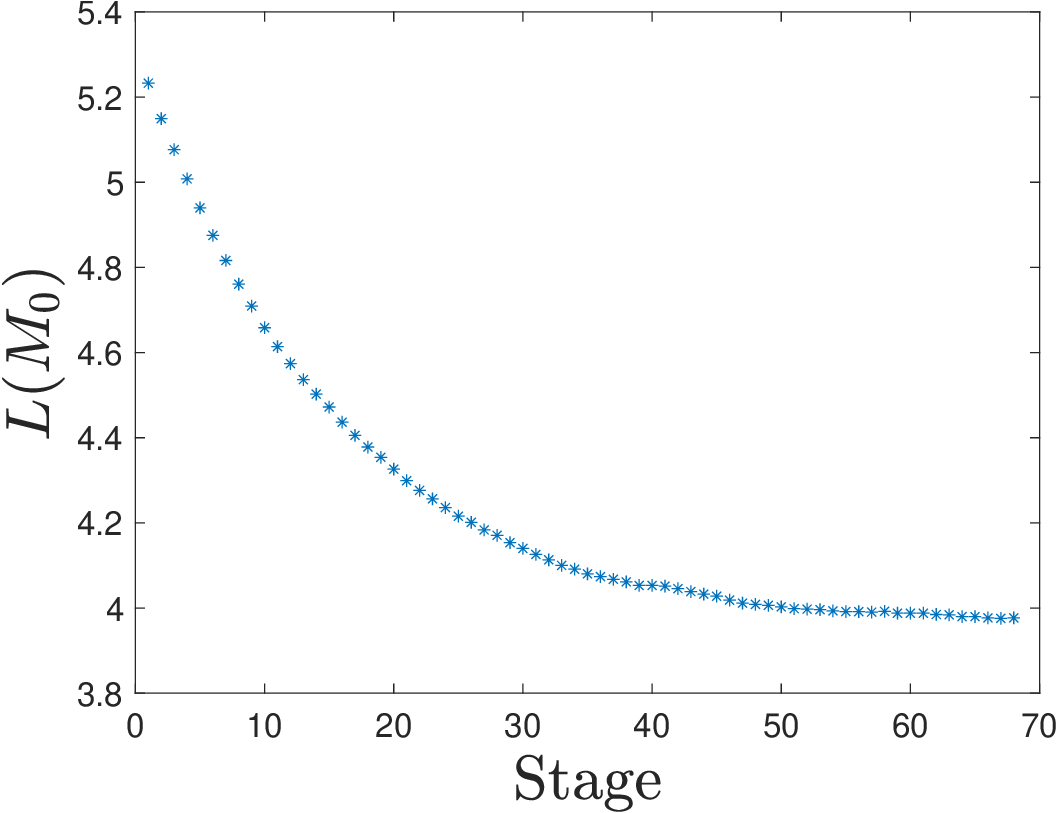}}
    \caption{The (a) mean outbreak duration, (b) mean final outbreak size, and (c) mean outbreak peak as a function of the stage number of Algorithm \ref{alg:functional} with $N_{\rm S} = 68$ stages, the graph $G^{\mathrm{RL}}$ that we generate {with} Algorithm \ref{alg:generating_G} (for $n_{\mathrm{WS}} = 12$, $N_{\mathrm{WS}} = 20$, and  $k_{\mathrm{WS}} = 6$), and $N_{\mathrm{sim}} = 1000$ simulations of SIR dynamics [for parameter configurations with $\alpha = 0.1$, $\delta_* = 0.2$, $\delta_{**} = 1$, $\beta_* = 0.125$, and the associated value of $\beta_{**}$ that we obtain using \eqref{eqn:compensation}]. In (d), {we show} the map function $L(M_0) = L(M_0, P_e(D_\delta^{(s)}, \mathbf{0}, t = 0.025))$ for the planted partition $M_0$ of $G^{\mathrm{RL}}$ as a function the stage number of Algorithm \ref{alg:functional} with the same inputs as in (a)--(c), where the node-absorption rates in the diagonal {entries} of $D_\delta^{(s)}$ correspond to
    the parameter configuration in stage $s$ of Algorithm \ref{alg:functional}. }
    \label{fig:FunctionalQuant}
\end{figure}

\subsubsection{Effective community structure of a single planted community} \label{next}

The 20 ring-lattice subgraphs $G_1,\ldots,G_{20}$ of the network $G^{\mathrm{RL}}$ in Section \ref{sec:network_SIR} are isomorphic and randomly connected according to the procedure in Algorithm \ref{alg:generating_G}. We now examine the effective community structure of one ring-lattice subgraph of $G^{\mathrm{RL}}$. We use Algorithm \ref{alg:info_abs_partB} with the input $P_e(D_\delta, \mathbf{0},t)$ for the adjacency matrix of $G^{\mathrm{RL}}$, {Markov times $t \in (0.01,0.05)$, and node-absorption rates $\delta_1,\ldots, \delta_{240}$ {from} {the parameter configurations in three stages of} Algorithm \ref{alg:functional}.} We use $H = \mathbf{0}$ as an input of Algorithm \ref{alg:info_abs_partB} because the influence of the absorption on the effective community structure is particularly noticeable when $H = \mathbf{0}$ [e.g., see Figures \ref{fig:4commMarkovTimes}(b,d)]. 

{We use the parameter configurations {from} the} following three stages of Algorithm \ref{alg:functional}: the initial stage (stage 1), the stage with the peak duration (stage 29), and the final stage (stage 68). We then select the planted community $G_5$ [which consists of the nodes with labels 49--60 in Figures \ref{fig:Stages}(c,d)]. We choose this planted community because it has a relatively large number of bridging nodes (these are {nodes} 50, 53, 54, 56, and 59) and because {the node-absorption rates of these nodes increases from $\delta_{*}$ to $\delta_{**}$ in one of the three examined stages.} Given a partition $M$ of $G^{\textnormal{RL}}$, we refer to a subset of the planted community $G_5$ as a ``subcommunity'' of $G_5$ if it is the intersection between {$G_5$ and a community in $M$}. 

In Figures \ref{fig:Stages}(a) and \ref{fig:Stages}(b){,} we show the number of communities of $G^{\mathrm{RL}}$ and number of subcommunities of $G_5$, respectively, as functions of the Markov time $t$ for partitions that we obtain using Algorithm \ref{alg:info_abs_partB} with the input $P_e(D_\delta, \mathbf{0},t)$, {where the node-absorption rates in the diagonal {entries} of $D_\delta$ {correspond to} the parameter configurations in the initial stage (see the dash-dotted green curve), the stage with the peak duration (see the solid red curve), and the final stage (see the dashed blue curve).} In Figures \ref{fig:Stages}(a,b), we observe that the number of communities of the network $G^{\mathrm{RL}}$ [in Figure \ref{fig:Stages}(a)] and the number of subcommunities of the subgraph $G_5$ [in Figure \ref{fig:Stages}(b)] increase with the stage number. 

In Figures \ref{fig:Stages}(c,d), we show the subcommunities of $G_5$ for Markov time $t = 0.025$. We select $t = 0.025$ because the curves in Figures \ref{fig:Stages}(a,b) are either flat near this value or change relatively little near it. In Figure \ref{fig:Stages}(c), we see that the nodes with the large node-absorption rate $\delta_{**}$ (i.e., nodes 54, 56, and 59) are in different communities but that the disease can flow through the blue community and enter a different planted community in the peak-duration stage. For example, the increase in the transmission {intensities} of some of the nodes that belong to the same community (e.g., nodes 50, 51, and 53) implies that the disease can still spread to other planted communities (through the bridging nodes 50 and 53), so the outbreak lasts longer in this stage than in the initial stage [see Figure \ref{fig:FunctionalQuant}(a)].

In Figure \ref{fig:Stages}(d), we see that there are more bridging nodes with the large node-absorption rate $\delta_{**}$ in the final stage than in the peak-duration stage. (Specifically, nodes 50 and 53 have the large node-absorption rate in the final stage.) Therefore, the spread of the disease to other planted communities is less likely in the final stage than in the peak-duration stage. Specifically, in stage 68 (unlike in stage 29), the disease can {disappear} more easily at nodes 50 and 53 (which belong to different communities in the final stage) {in the final stage} because nodes 50 and 53 have large node-absorption rates. Therefore, we expect that the outbreak duration, final outbreak size, and outbreak peak in Figures \ref{fig:FunctionalQuant}(a)--(c) in the final stage are smaller than in earlier stages.


\begin{figure}[H]
     \centering
     \subfloat[][\centering{Total number of communities as a function of the Markov time $t$}]{\includegraphics[scale = 0.45]{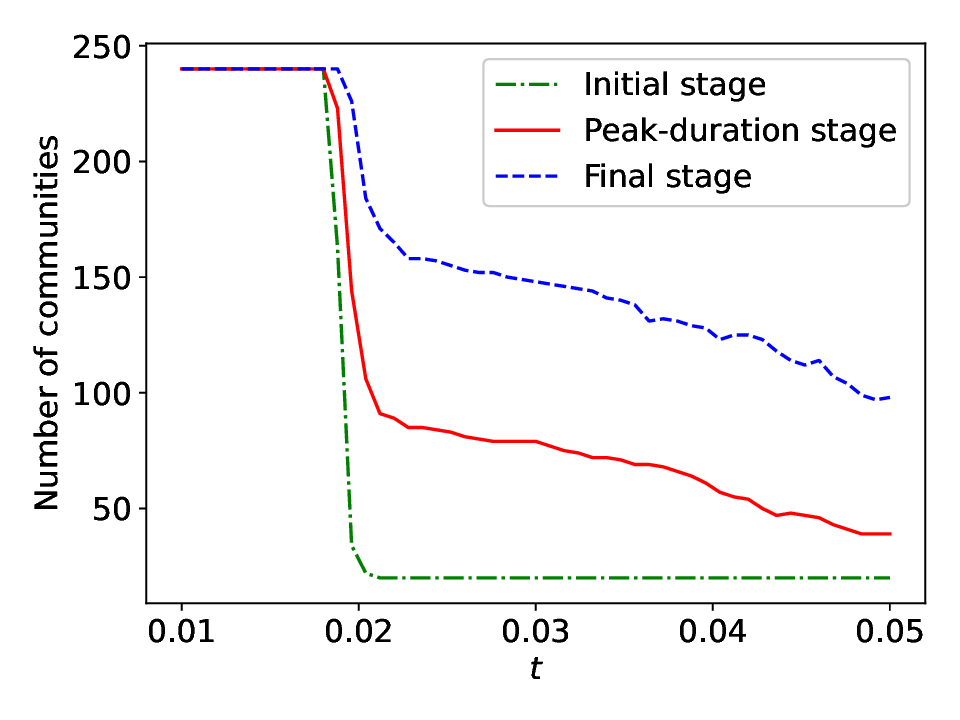}}
     \subfloat[][\centering{Number of subcommunities of the planted community $G_5$ as a function of the Markov time $t$}]{\includegraphics[scale = 0.45]{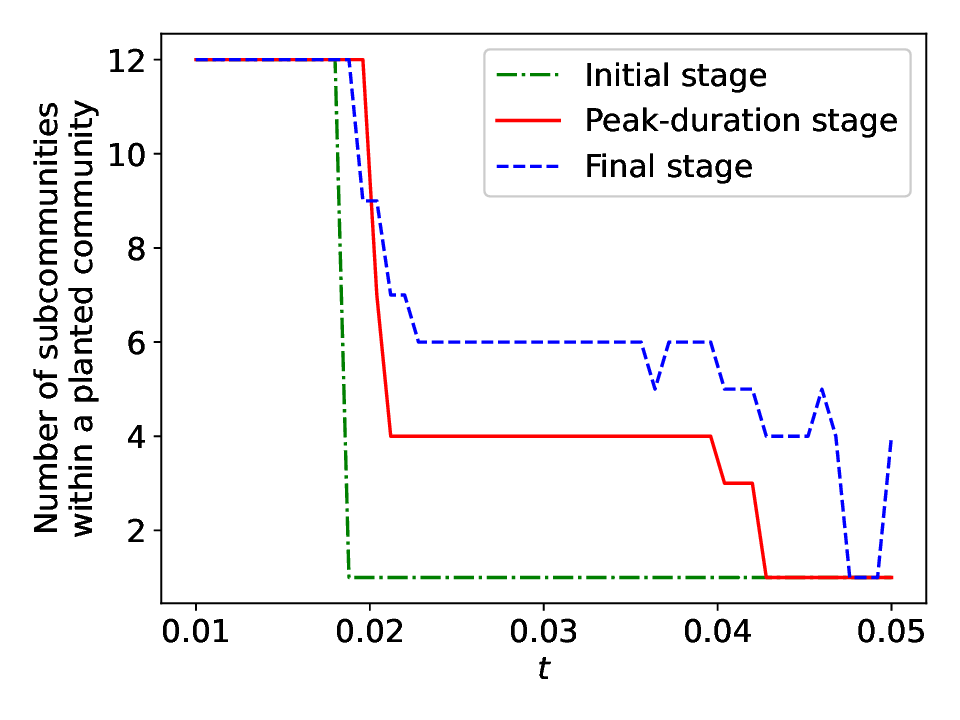}}\\
     \subfloat[][\centering{Peak-duration stage ($t = 0.025$), planted community $G_5$}]{\includegraphics[scale=0.7]{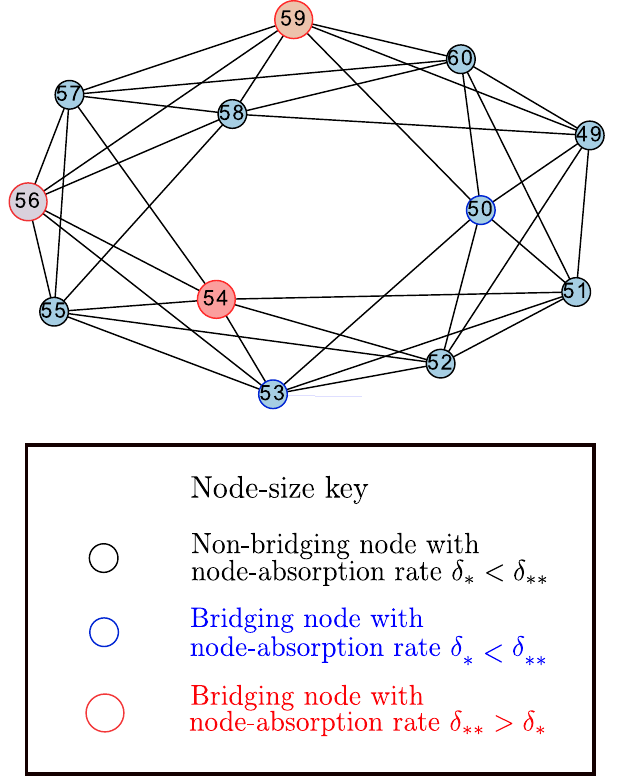}} \quad \quad 
     \subfloat[][\centering{Final stage ($t = 0.025$), planted community $G_5$}]{\includegraphics[scale=0.7]{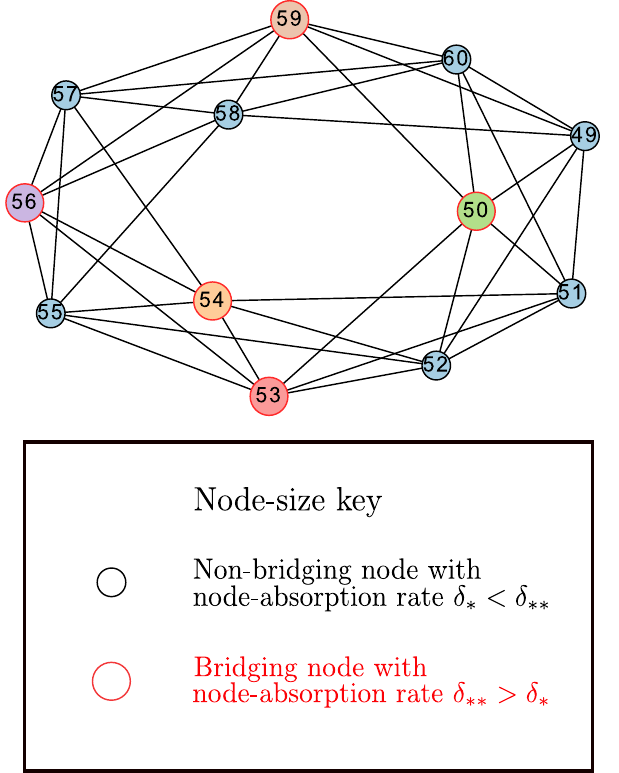}}
     \caption{Comparison of community structures in the initial, peak-duration, and final stages of Algorithm \ref{alg:functional} for SIR dynamics on $G^{\mathrm{RL}}$.
     (a) The number of communities in $G^{\mathrm{RL}}$ that we obtain using Algorithm \ref{alg:info_abs_partB} with the input $P_e(D_\delta, \mathbf{0},t)$ for {the node-absorption rates {from} the parameter configurations in three stages of Algorithm \ref{alg:functional}.} (b) The number of subcommunities in the planted community $G_5$ that we obtain using Algorithm \ref{alg:info_abs_partB} for {the node-absorption rates {from} the parameter configurations in three stages of Algorithm \ref{alg:functional}}. (c,d) The subcommunities of $G_5$ that we obtain using Algorithm \ref{alg:info_abs_partB} in (c) the peak-duration stage with $t = 0.025$ and (d) the final stage with $t = 0.025$. Each color indicates a subcommunity. 
     }
     \label{fig:Stages}
\end{figure}

\section{Conclusions and Discussion}\label{sec:discussion}

We adapted the community-detection algorithm InfoMap by using absorbing random walks, which are important for a variety of dynamical processes on networks. Through theoretical analysis and numerical computations of examples with small networks, we demonstrated that incorporating heterogeneous node-absorption rates leads to an effective community structure that can differ markedly from conventional notions of community structure. Through simulations of an SIR compartmental model of disease spread, we illustrated that such effective community structure can significantly influence dynamical processes on a network.

In our work, we considered a specific compartmental model on a specific type of random graph. It is important to examine the relationships between node-absorption rates, effective community structure, and dynamics in more general settings, including for other random-graph models and for empirical networks.
Some relevant settings include mobility networks that link spatial locations, with node-absorption rates that reflect variations in habitat quality; sexual networks, with nodes corresponding to individuals and node-absorption rates that reflect heterogeneities in treatment rates in different subsets of a population; and propagation of online content, with node-absorption rates that reflect different rates of content removal.  

The community-detection algorithm InfoMap is based on random walks, so it is natural to adapt it to absorbing random walks. However, there are numerous approaches to community detection \cite{fortunato2016community, Porter2009}, and it is worthwhile to adapt other approaches, such as modularity maximization \cite{newman2018networks} and statistical influence using stochastic block models \cite{peixoto2019bayesian}, to account for node-absorption rates. Absorption-scaled graphs provide a useful way to adapt community-detection methods (including ones that are not based on random walks) to account for heterogeneous node-absorption rates. Community structure depends not only on network structure but also on network dynamics (see, e.g., \cite{jeub2015}), and it is important to use a variety of perspectives to examine the ``effective community structure'' that is associated with different dynamical processes.

\appendix

\section{Appendix}\label{appendix}

We now prove Propositions \ref{prop:z0_z1}, \ref{prop:L0L1}, and \ref{prop:d0_d1} from Section \ref{sec:relating_G_delta}. We start with Proposition \ref{prop:z0_z1}.

\setcounter{proposition}{3}

\begin{proposition}
Let $P_0 = AW^{-1}$ be the transition-probability matrix of the discrete-time Markov chain that is associated with $\tilde{\mathcal{L}}(D_\delta,\mathbf{0})$, and let $P_1 = (A + D_\delta)(W + D_\delta)^{-1}$ be the transition-probability matrix of the discrete-time Markov chain that is associated with $\tilde{\mathcal{L}}(D_\delta, I)$. Let $\pi$ and $\pi'$ be the stationary distributions that are associated with $P_0$ and $P_1$, respectively. Let $Z_i$ be the fundamental matrix that is associated with $P_i$ (with $i \in \{1,2\}$). Let $U := (1/(\vec{\delta}^{\T}\vec{u}))\vec{u}\vec{1}^{\T}$ and $\alpha := \vec{\delta}^{\T}\vec{u}/(\vec{w}^{\T}\vec{u}+\vec{\delta}^{\T}\vec{u})$, where $\vec{u} = (u_1,\ldots,u_n)^{\T}$ is a vector in the kernel $\Ker {(W - A)}$ with positive entries {$u_i$} such that $\sum_{i=1}^n u_i = 1$. We have that
\begin{equation}\tag{29}
	Z_1 = W^{-1}(W+D_\delta)\left[Z_0 + \alpha(1-\alpha)\pi \vec{1}^{\T} - \alpha\left(Z_0D_\delta U + W\frac{\vec{u}\vec{\delta}^{\T}}{\vec{\delta}^{\T}\vec{u}}W^{-1}Z_0(I-\alpha D_\delta U)\right)\right] \, .
\end{equation}
\end{proposition}

\begin{proof}
Let $\vec{u}$ be a column vector in $\Ker {(W - A)}$ whose entries $u_i$ are all positive and sum to $1$. It follows that $\pi = W\vec{u}/(\vec{w}^{\T}\vec{u})$, $\pi' = (W+D_\delta)\vec{u}/\left(\vec{w}^{\T}\vec{u}+\vec{\delta}^{\T}\vec{u}\right)$, and 
\begin{equation}
	\pi' = (1-\alpha)\pi + \alpha \frac{D_\delta \vec{u}}{\vec{\delta}^{\T}\vec{u}} \, .
\end{equation}
Additionally,
\begin{equation*}
	Z_1 = \left(I - P_1 + \pi'\vec{1}^{\T}\right)^{-1}  \, ,   
\end{equation*}
which we can write as
\begin{equation} \label{eqn:pfZ1}
	Z_1 = W^{-1}(W+D_\delta) \left[I - P_0 + \pi \vec{1}^{\T} -\alpha\pi\vec{1}^{\T} + \alpha \frac{D_\delta\vec{u}}{\vec{\delta}^{\T}\vec{u}}\vec{1}^{\T} + \left((1-\alpha)\pi + \alpha \frac{D_\delta \vec{u}}{\vec{\delta}^{\T}\vec{u}}\right) \vec{1}^{\T}D_\delta W^{-1} \right]^{-1} \, .
\end{equation}

We now use the Sherman--Morrison formula to compute the inverse in (\ref{eqn:pfZ1}). The Sherman--Morrison formula states that if $F$ is a non-singular matrix and $\vec{v}_1$ and $\vec{v}_2$ are column vectors such that $1+{\vec{v}_2^{\T}F^{-1}\vec{v}_1} \neq 0$, then 
\begin{equation} \label{eqn:sherman}
	 \left(F + \vec{v}_1 \vec{v}_2^{\T}\right) ^{-1}=  F^{-1} - \frac{F^{-1}\vec{v}_1\vec{v}_2^{\T}F^{-1}}{1 + \vec{v}_2^{\T} F^{-1} \vec{v}_1}  \, .  
\end{equation}

Define 
\begin{align}
F_0 &:= I - P_0 + \pi \vec{1}^{\T} \,,\notag \\ 
F_1 &:= I - P_0 + \pi \vec{1}^{\T} -\alpha\pi\vec{1}^{\T} \,,\notag\\ 
F_2 &:= I - P_0 + \pi \vec{1}^{\T} -\alpha\pi\vec{1}^{\T} + \alpha \frac{D_\delta\vec{u}}{\vec{\delta}^{\T}\vec{u}}\vec{1}^{\T} \,,\notag \\  
F_3 &:= I - P_0 + \pi \vec{1}^{\T} -\alpha\pi\vec{1}^{\T} + \alpha \frac{D_\delta\vec{u}}{\vec{\delta}^{\T}\vec{u}}\vec{1}^{\T} + \left((1-\alpha)\pi + \alpha \frac{D_\delta \vec{u}}{\vec{\delta}^{\T}\vec{u}}\right) \vec{1}^{\T}D_\delta W^{-1} \,. 
\end{align}
{{Additionally, recall that the} fundamental matrix $Z$ of a regular Markov chain with stationary distribution $\vec{p}$ satisfies
\begin{equation} \label{eqn:one_zo_pi}
	 Z \vec{p} = \vec{p} \quad \text{ and } \quad \vec{1}^{\T}Z = \vec{1}^{\T}\,.
\end{equation}}
{Using} (\ref{eqn:sherman}) and (\ref{eqn:one_zo_pi}), we obtain 
\begin{align} \label{eqn:pfZ2}
	F_1^{-1} &= \left(F_0
 - \alpha\pi\vec{1}^{\T}\right)^{-1} =  Z_0 + \frac{\alpha}{1-\alpha}\pi\vec{1}^{\T} \, , \notag \\
	F_2^{-1} &= \left(F_1
 + \alpha\frac{D_\delta \vec{u}}{\vec{\delta}^{\T}\vec{u}}\vec{1}^{\T}\right)^{-1} =  Z_0 + \alpha\pi\vec{1}^{\T} -\alpha Z_0D_\delta U\, , \notag  \\
	 F_3^{-1} &= \left(F_2
 + \left((1-\alpha)\pi + \alpha \frac{D_\delta \vec{u}}{\vec{\delta}^{\T}\vec{u}}\right) \vec{1}^{\T}D_\delta W^{-1} \right)^{-1}  \notag \\
  &= Z_0 + \alpha(1-\alpha)\pi \vec{1}^{\T} - \alpha\left[Z_0D_\delta U +  W\frac{\vec{u}\vec{\delta}^{\T}}{\vec{\delta}^{\T}\vec{u}}W^{-1}Z_0(I-\alpha D_\delta U)\right] \, .    
\end{align} 
From equation \eqref{eqn:pfZ1}, it follows that $Z_1 = W^{-1}(W+D_\delta)F_3^{-1}$. Combining this relation with \eqref{eqn:pfZ2} yields \eqref{eqn:pfZ0}.
\end{proof}

We now prove Proposition \ref{prop:L0L1}.

\setcounter{proposition}{7}

\begin{proposition} 
Let $\tilde{\mathcal{L}}_1  := \tilde{\mathcal{L}}(D_\delta,I) = (W - A)(W + D_\delta)^{-1}$, and let $\vec{d}_1$ be the diagonal of $D_\delta(W + D_\delta)^{-1}$. Additionally, let $U:= \vec{u}\vec{1}^{\T}/(\vec{\delta}^{\T}\vec{u})$, $U_1 := (W + D_\delta)U$, and $D_1 := D_\delta(W+D_\delta)^{-1}$. We have that
\begin{equation} \tag{36}
	\tilde{\mathcal{L}}_1^{\vec{d}_1} = (W+D_\delta) \mathcal{L}^{\vec{\delta}}  
\end{equation}
and 
\begin{equation} \tag{37}
	(\mathcal{L}+D_\delta)^{-1} = (W+D_\delta)^{-1}\left(U_1 + (I + \tilde{\mathcal{L}}_1^{\vec{d}_1}D_1)^{-1}\tilde{\mathcal{L}_1}^{\vec{d}_1}\right) \,.  
\end{equation}
\end{proposition}

\begin{proof}
The adjacency matrix $A_1 = P_1 = (A + D_\delta)(W + D_\delta )$ has the associated unnormalized graph Laplacian matrix $\tilde{\mathcal{L}}_1 = (W - A)(W + D_\delta)^{-1}$.

From Proposition \ref{prop:z0_z1}, it follows that 
\begin{equation}\label{eqn:1pfL01}
	 Z_1 = W^{-1}(W+D_\delta)[Z_0 + \alpha R ] \,,
\end{equation}
where
\begin{equation*}
	R =  (1-\alpha)\pi \vec{1}^{\T} - Z_0D_\delta U - WUD_\delta W^{-1}Z_0(I - \alpha D_\delta U) \, .
\end{equation*}
Proposition \ref{prop:compute_abs_inv} then implies that
\begin{equation}\label{eqn:0pfL01}
	\tilde{\mathcal{L}}_1^{\vec{d}_1} = (I - U_1D_1)Z_1(I - D_1U_1)\,.
\end{equation}
Additionally,
\begin{align}\label{eqn:23pfL01}
	I - U_1D_1 &= (W+D_\delta)(I-UD_\delta )(W+D_\delta)^{-1} \,,  \notag \\ 
	 I - D_1U_1 &= I-UD_\delta \,.
\end{align}

Substituting (\ref{eqn:23pfL01}) into (\ref{eqn:0pfL01}) yields 
\begin{equation}\label{eqn:4pfL0L1}
	\tilde{\mathcal{L}}_1^{\vec{d}_1} = (W+D_\delta)\mathcal{L}^{\vec{\delta}} + \alpha (W+D_\delta)(I-UD_\delta)W^{-1}R(I-D_\delta U) \,.
\end{equation}
Using the relations $\pi\vec{1}^{\T}(I-D_\delta U) = \mathbf{0}$, $D_\delta U(I-D_\delta U) = \mathbf{0}$, and $(I-\alpha D_\delta U)(I-D_\delta U) = I-D_\delta U$ yields
$R(I - D_\delta U)= -WUD_\delta W^{-1}Z_0(I-D_\delta U)$. We then use the fact that $(I-UD_\delta) UD_\delta =\mathbf{0}$ to obtain 
\begin{equation}\label{eqn:4pfL0L1_0}
	(I-UD_\delta)W^{-1}R(I-D_\delta U) = \mathbf{0} \, . 
\end{equation}	
Substituting (\ref{eqn:4pfL0L1_0}) into the right-hand side of (\ref{eqn:4pfL0L1}) yields (\ref{eqn:L0L1}).

We express the fundamental matrix $(\mathcal{L} + D_\delta)^{-1}$ as 
\begin{align}\label{eqn:5pfL01}
	(\mathcal{L} +D_\delta)^{-1} &= (W + D_\delta)^{-1}\left((W - A)(W + D_\delta)^{-1} + D_\delta(W + D_\delta)^{-1}\right)^{-1} \notag \\    
					&= (W + D_\delta) (\tilde{\mathcal{L}}_1 + D_1)^{-1} \, . 
\end{align}					
By Proposition \ref{prop:fund_and_absinv} and (\ref{eqn:L0L1}), we have 
 \begin{equation}\label{eqn:6pfL01}
 	(\tilde{\mathcal{L}}_1 + D_1)^{-1} = U_1 + (I + \tilde{\mathcal{L}}_1^{\vec{d}_1}D_1)^{-1}\tilde{\mathcal{L}_1}^{\vec{d}_1}  \,.
\end{equation}
Equations (\ref{eqn:5pfL01}) and (\ref{eqn:6pfL01}) then yield (\ref{eqn:L1fund}).

\end{proof}


We now prove Proposition \ref{prop:d0_d1}.

\begin{proposition} 
Let $\vec{d}' := \vec{d}_{\mathrm{s}}(D_\delta,I) = \vec{w} + \vec{\delta} =  (\omega_1+\delta_1,\ldots, \omega_n + \delta_n)^{\T}$ be the scaled rate vector that is associated with the absorption-scaled graph $\tilde{G}(D_\delta, I)$. With $\mathcal{L} = W - A$, $\alpha := \vec{\delta}^{\T}\vec{u}/(\vec{w}^{\T}\vec{u} + \vec{\delta}^{\T}\vec{u})$, $\pi = W\vec{u}/(\vec{w}^{\T}\vec{u})$, $Z_0 = \left(I - AW^{-1} + \pi \vec{1}^{\T}\right)^{-1}$, and $Z_* = W^{-1}\left(Z_0 - \pi \vec{1}^{\T} \right)$, it follows that
\begin{equation} \tag{38}
	\mathcal{L}^{\vec{d}'} = \alpha^2\mathcal{L}^{\vec{\delta}} + \alpha(1-\alpha)\left(\mathcal{L}^{\vec{\delta}}\mathcal{L}Z_*+Z_*\mathcal{L}\mathcal{L}^{\vec{\delta}} \right) +(1-\alpha)^2Z_*   \, . 
\end{equation}
\end{proposition}

\begin{proof}
By Proposition \ref{prop:compute_abs_inv}, we have
\begin{equation}\label{eqn:1pf}
	\mathcal{L}^{\vec{d}'} = \left(I - \frac{\vec{u} \vec{1}^{\T}}{\vec{w}^{\T}\vec{u} + \vec{\delta}^{\T}\vec{u}}(W+D_\delta ) \right) W^{-1} Z_0 \left(I - \frac{1}{\vec{w}^{\T}\vec{u} + \vec{\delta}^{\T}\vec{u}}(W + D_\delta ) \vec{u}\vec{1}^{\T} \right)\, . 
\end{equation}

Furthermore, 
\begin{align}\label{eqn:23pf}
 	I - \frac{\vec{u} \vec{1}^{\T}}{\vec{w}^{\T}\vec{u} + \vec{\delta}^{\T}\vec{u}}(W + D_\delta )  &= \alpha \left(I - \frac{\vec{u}\vec{1}^{\T}}{\delta^{\T}\vec{u}}D_\delta\right) + (1 - \alpha)W^{-1}(I - \pi\vec{1}^{\T})W \, ,\notag \\
 	I - \frac{1}{\vec{w}^{\T}\vec{u} + \vec{\delta}^{\T}\vec{u}}(W + D_\delta ) \vec{u}\vec{1}^{\T}  &= \alpha\left(I - D_\delta\frac{\vec{u}\vec{1}^{\T}}{\delta^{\T}\vec{u}}\right) + (1 - \alpha)(I - \pi\vec{1}^{\T}) \,.
\end{align}

Substituting (\ref{eqn:23pf}) into (\ref{eqn:1pf}) yields
\begin{align} \label{eqn:4pf}
	 \mathcal{L}^{\vec{d}'} &= \alpha^2\left(I - \frac{\vec{u}\vec{1}^{\T}}{\delta^{\T}\vec{u}}D_\delta\right)W^{-1}Z_0\left(I - D_\delta\frac{\vec{u}\vec{1}^{\T}}{\delta^{\T}\vec{u}}\right) +  \alpha(1-\alpha)\left(I - \frac{\vec{u}\vec{1}^{\T}}{\delta^{\T}\vec{u}}D_\delta\right)W^{-1}Z_0(I-\pi\vec{1}^{\T}) \notag \\
	&\quad + \alpha(1-\alpha)  W^{-1}(I - \pi\vec{1}^{\T})W W^{-1} Z_0\left(I - D_\delta\frac{\vec{u}\vec{1}^{\T}}{\delta^{\T}\vec{u}}\right) + (1-\alpha)^2  W^{-1}(I - \pi\vec{1}^{\T})W W^{-1} Z_0 (I-\pi\vec{1}^{\T})\,. 
\end{align}

By Proposition \ref{prop:fund_and_absinv}, the first term of the right-hand side of (\ref{eqn:4pf}) is {$\alpha^2\mathcal{L}^{\vec{\delta}}$}. Note that $\pi$ is the stationary distribution of the Markov chain with transition-probability matrix $P_0$. Additionally, $Z_0$ is the fundamental matrix of this Markov chain. Therefore, using
(\ref{eqn:one_zo_pi}), it follows that
$W^{-1} Z_0 (I-\pi\vec{1}^{\T}) = W^{-1} (I-\pi\vec{1}^{\T})Z_0 = W^{-1}(Z_0-\pi\vec{1}^{\T}) = Z_*$ and  $ W^{-1}(I - \pi\vec{1}^{\T}) Z_0 (I-\pi\vec{1}^{\T}) =  W^{-1}(Z_0-\pi\vec{1}^{\T}) = Z_*$. Consequently, from (\ref{eqn:4pf}), we obtain 
\begin{equation} \label{eqn:5pf}
	 \mathcal{L}^{\vec{d}'} = \alpha^2 \mathcal{L}^{\vec{\delta}} +  \alpha(1-\alpha) \left[(I - \frac{\vec{u}\vec{1}^{\T}}{\delta^{\T}\vec{u}}D_\delta)Z_* +  Z_* \left(I - D_\delta\frac{\vec{u}\vec{1}^{\T}}{\delta^{\T}\vec{u}}\right)\right] + (1-\alpha)^2 Z_* \,.
 \end{equation}
Additionally, $\mathcal{L}^{\vec{\delta}}\mathcal{L} = I - \frac{\vec{u}\vec{1}^{\T}}{\delta^{\T}\vec{u}}D_\delta $ and $\mathcal{L}\mathcal{L}^{\vec{\delta}} = I - D_\delta\frac{\vec{u}\vec{1}^{\T}}{\delta^{\T}\vec{u}} $ (see Theorem 1 and Lemma 1 in \cite{jacobsen2018generalized}), so (\ref{eqn:0pf}) follows from (\ref{eqn:5pf}).
\end{proof}


\section*{Data and code availability}

The code that yields the numerical results in Figures \ref{fig:FunctionalQuant} and \ref{fig:Stages} of Section \ref{sec:small_world} are available at \url{https://gitlab.com/esteban_vargas_bernal/extending-infomap-to-absorbing-random-walks}.

\section*{Acknowledgements}

We thank the National Science Foundation (through grants DMS 1814737 and DMS 1440386) and the Fulbright International Program for financial support.  


\bibliography{biblio_v11}
\bibliographystyle{plain}

\end{document}